\documentstyle[preprint,prb,aps,oldlfont]{revtex} 
\begin{document}
\draft
\date{\today}
\title{%
Charge and Spin Gap Formation\protect\\
in Exactly Solvable Hubbard Chains with
Long-Range Hopping}

\author{Florian Gebhard and Andreas Girndt}
\address{Dept.~of Physics and Materials Sciences Center, Philipps University,
D-35032~Marburg, Germany}

\author{Andrei E. Ruckenstein}
\address{Serin Physics Laboratory, Rutgers University, P.O.~Box 849,
Piscataway, NJ 08855-0849}
\maketitle%
\begin{abstract}%
We discuss the transition from a metal to charge or spin insulating
phases characterized by the opening of a gap in the charge or spin
excitation spectra, respectively.
These transitions are addressed within
the context of two exactly solvable Hubbard and tJ
chains with long range, $1/r$ hopping.
We discuss the specific heat, compressibility, and magnetic
susceptibility of these models as a function of
temperature, band filling, and interaction strength.
We then use conformal field theory techniques to
extract ground state correlation functions.
Finally, by employing the $g$-ology analysis we show that
the charge insulator transition is accompanied by an infinite
discontinuity in the Drude weight of the electrical conductivity.
While the magnetic properties of these models
reflect the genuine features of strongly correlated electron systems,
the charge transport properties, especially near the Mott-Hubbard transition,
display a non-generic behavior.
\end{abstract}

\narrowtext
\section{Introduction}
Recently~\cite{prl} we introduced a Hubbard-like model which
describes spin-$1/2$ Fermions on a chain of $L$ sites, hopping with
long-range amplitude, $t_{l,m} = it (-1) ^{l-m} \left[d(l-m)\right]^{-1}$.
Here, $d(l-m) = (L/\pi )\sin [\pi (l-m)/L]$ is the chord distance between
sites $l$ and $m$ on the chain closed into a ring
(the lattice spacing~$a$ is set to unity).
In the thermodynamic limit, $L\to \infty$, and for a fixed distance,
$(l-m)$, the purely imaginary hopping
$t_{l,m} = t_{m,l}^{*}$ becomes
$t_{l,m} \to it/(l-m)$, leading to the Hubbard Hamiltonian with
``$1/r$-hopping'',
\begin{equation}
\hat{H} = \hat{T} + U \hat{D} =
\sum_{l \neq m = 1,\sigma}^{L} t_{l,m} \hat{c}_{l,\sigma}^{+}
\hat{c}_{m,\sigma}^{\mbox{}}
+ U \sum_{l=1}^{L} \hat{n}_{l,\uparrow}\hat{n}_{l,\downarrow} \quad .
\label{hamilt}
\end{equation}
Here $U$ is the strength of the usual
local Hubbard~\cite{Hubbard}
interaction.
For even~$L$ we choose antiperiodic boundary conditions, so that
the dispersion relation is linear in wave vector, namely, $\epsilon (k) = t k$
for $k=\Delta (m+1/2)$ ($\Delta =2\pi /L$,
$m=-L/2,\ldots\ ,L/2 -1)$. For $U=0$ the Fermi sea is the
ground state with all $k$-states from $k=-\pi$
to $k_F^e=\pi (n-1)$ filled~\cite{fermibody},
where $n = (N_{\uparrow}
+N_{\downarrow})/L$ is the total particle density (the ``filling'').
Note that the kinetic and potential energy operators are
respectively odd and even under partity, and thus parity
is, in general, not a good quantum number in our model.

In the large $U$ limit the Hamiltonian~(\ref{hamilt})
becomes the $1/r$-tJ model including
``pair-hopping" terms ($\hat{P}_{D=0}$ projects onto the subspace with
no double occupancies)~\cite{HarrisLange}:
\begin{eqnarray}
\hat{H}_{\rm tJ} & = &  \hat{P}_{D=0} \biggl\{
\hat{T} + \sum_{l\neq m} \frac{2 |t_{l,m}|^{2}}{U}
\left[ \hat{{\rm\bf S}}_{l} \cdot \hat{{\rm\bf S}}_{m} - \frac{1}{4}
\hat{n}_{l} \hat{n}_{m} \right] \nonumber \\
 & & \phantom{  \hat{P}_{D=0} \biggl\{ }
- \sum_{l \neq n \neq m \neq l} \frac{t_{l,n}t_{n,m}}{U}
\sum_{\sigma\sigma'} \left( \sigma\sigma' \right)
\hat{c}_{l\sigma}^{+}\hat{c}_{n-\sigma}^{+}
\hat{c}_{n-\sigma'}^{\mbox{}}\hat{c}_{m\sigma'}^{\mbox{}}
\biggr\} \hat{P}_{D=0} \quad .
\label{tJ}
\end{eqnarray}
As usual, $\hat{n}_{l} = \hat{n}_{l,\uparrow} +\hat{n}_{l,\downarrow}$,
and $\hat{{\rm\bf S}}_{l}$ is the spin-1/2 vector operator
($\hat{S}_{l}^{+}= \hat{c}_{l,\uparrow}^{+}
\hat{c}_{l,\downarrow}^{\phantom{+}}$,
$\hat{S}_{l}^{-}= \hat{c}_{l,\downarrow}^{+}
\hat{c}_{l,\uparrow}^{\phantom{+}}$,
$\hat{S}_{l}^{z}= (\hat{n}_{l,\uparrow} - \hat{n}_{l,\downarrow})/2$).
At half filling, where the first and third terms
in (\ref{tJ}) vanish, the model~(\ref{tJ})
reduces to the (parity symmetric) $(1/r)^2$-Heisenberg model introduced by
Haldane and Shastry~\cite{fdmh-bss}, with $J=4t^2 /U$.

In Ref.~\onlinecite{prl} we conjectured the full excitation
spectrum and associated degeneracies, and were thus able to calculate
the free energy. The exact solution allowed us to identify
two zero temperature ($T=0$) phase transitions: the first was
a Mott-Hubbard~\cite{Hubbard,Mott} metal-to-charge-insulator
transition (MCIT) in the half-filled $1/r$-Hubbard model. This was
signaled by the opening of a charge gap
for $U > U_{c} = W$ where $W=2\pi t$ is the electron bandwidth.
The second, a metal-to-spin insulator transition (MSIT) was associated
with the opening of a
spin gap in the $1/r$-tJ model~(\ref{tJ}) for
$J>J_{c} = 2W/\left[(1-n)\pi^{2}\right]$.

In this paper we clarify the nature of these
transitions and attempt to separate the features due to the
special form of the dispersion
in~(\ref{hamilt}) from more generic properties which
may be expected to survive for other dispersions and/or
in higher dimensions.
The plan of the paper is as follows: we start in section~\ref{sec2}
by recalling the form of our exact solution for the spectrum and
ground state energies, from which
we identify the location of the MCIT and MSIT.
In section~\ref{sec3} we discuss the thermodynamic properties of our
two models, namely the specific heat,
compressibility, and magnetic susceptibility.
In sec.~\ref{CFT} we use conformal field
theory techniques to extract
the long-range behavior of ground state correlation functions.
Sec.~\ref{g-ology} contains a discussion of the connection with
the ``g-ology" approach to one-dimensional systems,
which we exemplify by calculating the
Drude weight in the frequency dependent electrical conductivity
for the $1/r$-Hubbard model.
Finally, a summary and conclusions are presented in sec.~\ref{summary}.

\section{Ground state properties and excitation spectra}
\label{sec2}
We recall the effective Hamiltonian, introduced in ref.~\onlinecite{prl},
\begin{equation}
\hat{H}^{\rm eff} =
\sum_{-\pi < {\cal K} < \pi} \left\{
 \sum_{\sigma} h_{{\cal K},\sigma}^{s} \widetilde{n}_{{\cal K},\sigma}^{s}
+ h_{{\cal K}}^{d} \widetilde{n}_{{\cal K}}^{d}
+ h_{{\cal K}}^{e} \widetilde{n}_{{\cal K}}^{e}
+ J_{{\cal K}} \left[
\widetilde{n}_{{\cal K}-\Delta}^{d}\widetilde{n}_{{\cal K}}^{e} -
\widetilde{n}_{{\cal K}-\Delta,\uparrow}^{s}
\widetilde{n}_{{\cal K},\downarrow}^{s}
\right]
\right\}
\label{heff}
\end{equation}
which describes spin ($\widetilde{s}_{{\cal K}\sigma}$;
$S_{\cal K}^z = \pm 1/2$, $C_{\cal K}=0$)
and charge ($\widetilde{d}_{\cal K},\widetilde{e}_{\cal K}$;
$S_{\cal K}=0$, $C_{\cal K}^z = \pm 1/2$) degrees of freedom
in an occupation number representation with a hard core constraint,
$\sum_{\sigma} \widetilde{n}_{{\cal K},\sigma}^{s}
 + \widetilde{n}_{{\cal K}}^{d} + \widetilde{n}_{{\cal K}}^{e} = 1$
for each~${\cal K}$.
Furthermore, $h_{{\cal K},\sigma}^{s}=(t{\cal K})/2 - \mu_{\sigma}$,
$h_{{\cal K}}^{d}=-(t{\cal K})/2 - 2\mu +U$,
$h_{{\cal K}}^{e}=-(t{\cal K})/2$,
and $J_{{\cal K}} = \left[ t(2{\cal K}-\Delta) - U +
\sqrt{(2\pi t)^{2} + U^{2} - 2tU(2{\cal K}-\Delta)}\, \right]/2 \geq 0$.
The chemical potential in the presence of an external magnetic
field is given by
$\mu_{\sigma} = \mu - \sigma (g \mu_{B} {\cal H}_{0})/2$.
In the following we will set $\mu_{B} \equiv 1$, and $g=2$.
To be precise, we also restrict ourselves to
$t \geq 0$ and $U\geq - 2\pi t$ in which case
we were allowed to identify ${\cal K}_{\rm min}-\Delta \equiv
{\cal K}_{\rm max}$ because
$J_{{\cal K}}$ vanishes
for ${\cal K} = {\cal K} _{\rm min} = -\Delta (L-1)/2$.

Note that, formally, the entire spectrum {\em appears} to
display spin-charge separation {\em at all energies} in the sense
that the Hamiltonian splits up
into independent spin and charge contributions (``strong'' spin-charge
separation).
In reality, however, spin and charge excitations are coupled
by the constraint,
$\sum_{\sigma} \widetilde{n}_{{\cal K},\sigma}^{s}
 + \widetilde{n}_{{\cal K}}^{d} + \widetilde{n}_{{\cal K}}^{e} = 1$;
and spin-charge separation only
occurs at sufficiently low energies/temperatures,
where spin and charge excitations
contribute independently to various physical properties.

As already discussed above, we will
concentrate on the physics of two special limits:
(i)~the $1/r$-Hubbard model in the vicinity of half filling,
$n \lesssim 1$, and
(ii)~the $1/r$-tJ model  for $n < 1$.
The latter is obtained
by taking the limit
$U \to \infty$ of~(\ref{heff}) which projects out all double occupancies
($h_{{\cal K}}^{d} \to \infty$).
With the help of the completeness constraint
one arrives at the tJ effective Hamiltonian,
\begin{equation}
\hat{H}_{\rm tJ}^{\rm eff} =
\hat{H}_{\rm t}^{\rm eff} + \hat{H}_{\rm J}^{\rm eff}
=  \sum_{- \pi < {\cal K} < \pi}
\left[ (-t{\cal K}) \widetilde{n}_{{\cal K}}^{e}
- J_{{\cal K}}
\widetilde{n}_{{\cal K}-\Delta,\uparrow}^{s}
\widetilde{n}_{{\cal K},\downarrow}^{s}
\right]
\label{tJeff},
\end{equation}
where, for $J=4t^2/U \ll 1$,
the exchange coupling,  $J_{\cal K}(U/t)$, reduces
to $J_{\cal K} = (J/4) \left[ \pi^2 -
\left( {\cal K} -\Delta/2\right)^2\right]$.
Below, we will adopt the usual
standpoint for the tJ model~\cite{AndZou},
and treat~$J$ as an {\em independent} parameter.

\subsection{1/r-Hubbard Model}
\label{HMsec}
{}From eq.~(\ref{heff}) we can immediately extract the form of
eigenstates in an occupation number representation
in terms of the effective spin ($\widetilde{s}_{\cal K,\sigma}$)
and charge ($\widetilde{e}_{\cal K} ,\widetilde{d} _{\cal K}$)
degrees of freedom.
For example, the ground state is expressed symbolically as
\[
\text{ground state:}\quad
\left[ \uparrow \downarrow \right] \ldots \left[ \uparrow \downarrow \right]
\biggr|_{{\cal K}_{F}=\pi (2n-1)} \circ \ldots \circ
\]
where $\uparrow$, $\downarrow$, $\bullet$, and $\circ$ represent
$\widetilde{s}_{{\cal K},\uparrow}$, $\widetilde{s}_{{\cal K},\downarrow}$,
$\widetilde{d}_{{\cal K}}$, $\widetilde{e}_{{\cal K}}$, respectively.
We brace those pairs at ${\cal K} - \Delta$ and ${\cal K}$
which contribute an interaction $J_{{\cal K}}$ {}~\cite{singuletts}.
The ground state can be regarded as a short-range RVB state
in ${\cal K}$-space obtained by filling ${\cal K}$ states
with $\left[ \uparrow \downarrow \right]$-pairs from
${\cal K} = -\pi + \Delta/2$ to ${\cal K} = {\cal K}_F - \Delta/2$,
where ${\cal K}_{F} = \pi (2n -1)$ .
Although the real-space structure
of the ground state wave function is not immediately apparent,
it should be clear, however,
that it also consists of long-range (overlapping) singlet pairs.
In fact, the recently constructed wave functions for the~$U=\infty$ limit
(the ``$1/r$-t-model'')~\cite{WangColeman} are genuine
RVB states~\cite{AndersonRVB} of the
Gutzwiller-Jastrow type.

The corresponding ground state energy density, $e_{0} (n)$,
can easily be obtained by applying the effective Hamiltonian~(\ref{heff})
to the symbolic ground state wave function given above:
$e_0(n) =  (1/2\pi) \int_{-\pi}^{{\cal K}_F} d{\cal K}
\left[ J({\cal K})/2 \right] +
(1/2\pi)\int_{{\cal K}_F}^{\pi} d{\cal K} (-t{\cal K})$,
where the factor $1/2$ in the first term
takes into account the fact that only every second
${\cal K}$ value contributes to the first integral. A simple calculation
then gives
\begin{equation}
e_{0} (n \leq 1) = \frac{Un - W(1-n)n}{4}
- \frac{1}{24WU} \left[
\left( W+U \right)^{3} - \left( \left( W+U \right)^{2} - 4WUn
\right)^{3/2} \right] ,
\label{groundstate}
\end{equation}
where $W =2\pi t$ is the bandwidth.
Using particle-hole symmetry~\cite{howtodo1} we
obtain $e_{0}(n\geq 1) = e_{0} (2-n) + U(n-1)$, and correspondingly,
for the chemical potential at zero temperature,
$\mu (n < 1) =  \partial e_{0}(n<1) / \partial n$,
$\mu (n >1) = U - \mu (2-n)$. From~(\ref{groundstate}) one can easily see that
$\Delta \mu_c \equiv \mu_{+} (n=1) - \mu _{-} (n=1) =
(\lim _{n\rightarrow 1^{+}} - \lim _{n\rightarrow 1^{-}} )\mu (n)$
becomes finite for $U>U_{c} = W$: $\Delta \mu_c = U - U_c$.
Equivalently, the charge compressibility,
$\kappa =\partial n /\partial \mu$ at $n=1$ vanishes for $U$ above $U_c$,
as expected at a Mott metal-to-charge-insulator transition (MCIT).

We are now in the position to identify the low-lying spin and
charge excitations which we discuss separately
for the half-filled and less than half-filled cases:

\subsubsection{Half-filling: $n=1$}
We begin with the
{\em spin excitations}.
The four degenerate $S=0$, $S=1$ lowest-lying
spin excitations are represented by
\[
\text{spin-exc:\quad}
\left[ \uparrow \downarrow \right] \ldots \left[ \uparrow \downarrow \right]
\sigma\biggr|_{{\cal K}_1}
\left[ \uparrow \downarrow \right] \ldots \left[ \uparrow \downarrow \right]
\sigma'\biggr|_{{\cal K}_{2}}
\left[ \uparrow \downarrow \right] \ldots \left[ \uparrow \downarrow \right]
\quad .
\]
For ${\cal K}_2 = {\cal K}_1 + \Delta$ there is a triplet excitation only.
In the thermodynamic limit the excitation energy is given by
$\delta E_S ({\cal K}_1, {\cal K}_2) = \left[ J({\cal K}_1) +
J ({\cal K}_2) \right]/2$,
and we may thus identify the lowest-lying spin excitation with two
{\em spinons}. They are characterized as spin-1/2 objects which always come
in pairs, and are separated in ${\cal K}$-space by an {\em even multiple}
of $\Delta$.
We thus rescale ${\cal K} = 2 {\cal K}^{\prime}$ to retain the proper spacing
$\Delta= 2 \pi/L$ of ${\cal K}^{\prime}$ values.
The spinons then have the dispersion relation,
\begin{equation}
\epsilon_{s} ({\cal K}) =
J(2{\cal K})/2 = \left( \sqrt{W^{2} + U^{2} - 4 W U {\cal K}/\pi}
+ 2 W {\cal K}/\pi - U \right)/4 \geq 0 \quad ; \quad
|{\cal K}| < \pi/2 \, ,
\label{spinondispersion}
\end{equation}
which we depict in fig.~\ref{fig1}.
The spin excitations are always gapless at ${\cal K} = - \pi/2$, and
at ${\cal K} = \pi/2$ for $U/W> 1$. Their corresponding
velocities
at ${\cal K} = \pm \pi/2$ can now be calculated from $v_s ({\cal K}) =
\partial\epsilon_s ({\cal K})/ \partial {\cal K}$ as
\begin{mathletters}%
\label{spinonvelocities}%
\begin{eqnarray}
v_s^R & \equiv & v_s ({\cal K} = -\pi/2)
= \frac{v_F}{U/W+1} \quad \text{for all} \quad U/W
\label{vsr}\\[3pt]
v_s^L & \equiv &  v_s ({\cal K} = \pi/2)
= - \frac{v_F}{U/W-1} \quad \text{for} \quad U/W > 1
\quad . \label{vsl}
\end{eqnarray}%
\end{mathletters}%
Here, $v_F = t$ is the Fermi velocity of the bare particles.
The density of states for {\em spin excitations} is then calculated
as
$D_s (E) = (1/L) \sum_{-\pi/2<{\cal K}<\pi/2}
\delta\left(E-\epsilon_s({\cal K})\right)
= 1/(2\pi) \int_{-\pi/2}^{\pi/2}
d{\cal K}\, \delta\left(E-\epsilon_s({\cal K}) \right)$.
With the help of eqs.~(\ref{spinonvelocities}), for
low energies we obtain
\begin{equation}
D_s (E \to 0) =
\left\{ \begin{array}{cll}
1/\left( 2\pi v_{s}^{R}\right) &\text{for}& U/W < 1\\[3pt]
 1/\left( 2\pi v_{s}^{R}\right) + 1/\left( 2\pi |v_{s}^{L}|\right) &
\text{for} & U/W> 1
\end{array}
\right.
\quad .
\label{DsDOS}
\end{equation}

We next turn to the
{\em charge excitations}.
We restrict ourselves to a fixed particle number
($C^z =0$) and we only obtain two
of the four $C=0$, $C=1$ lowest-lying states. These
charge excitations are represented by
\[
\text{charge-exc:\quad}
\left[ \uparrow \downarrow \right] \ldots \left[ \uparrow \downarrow \right]
\circ\biggr|_{{\cal K}_{1}}
\left[ \uparrow \downarrow \right] \ldots \left[ \uparrow \downarrow \right]
\bullet\biggr|_{{\cal K}_{2}}
\left[ \uparrow \downarrow \right] \ldots \left[ \uparrow \downarrow \right]
\quad .
\]
For ${\cal K}_2 = {\cal K}_1 + \Delta$ there is a low lying
triplet excitation
($C=1$, $C^z=0$),
with the singlet excitation ($C=0$, $C^z=0$)
always at high energy (anti-bound state).
Again, we may identify the lowest-lying charge excitation with two
{\em chargeons}.
In the half-filled case we always get
a pair of a {\em holon} and a {\em doublon}
which are separated by an even multiple
of $\Delta= 2 \pi/L$. Rescaling ${\cal K}=2{\cal K}^{\prime}$
leads to the chargeon dispersion relation,
\begin{equation}
\epsilon_{c} ({\cal K}) =
 \left( \sqrt{W^{2} + U^{2} - 4 W U {\cal K}/\pi}
- 2 W {\cal K}/\pi + U \right)/4 \geq 0 \quad ; \quad
|{\cal K}| < \pi/2 \quad ,
\label{chargeondispersion}
\end{equation}
as depicted in fig.~\ref{fig1}.
The charge excitations are gapless at ${\cal K} = \pi/2$ for $U/W< 1$,
with a velocity given by
\begin{equation}
v_c^L  =  - \frac{v_F}{1-U/W} \quad \text{for} \quad U/W < 1\quad .
\label{vcl}
\end{equation}
For $U>W$ a gap,
$ 2 \epsilon_c (\pi/2) = U-W \equiv \Delta \mu_c$, opens in the charge
spectrum.
Correspondingly, the density of states for {\em charge excitations} at
low energies takes the form
\begin{equation}
D_c (E \to 0) =
\left\{ \begin{array}{cll}
1/ \left(2\pi |v_{c}^{L}|\right) &\text{for}& U/W < 1\\[3pt]
{\cal O}\left[ \exp \left(-E/\Delta\mu_c\right)\right]
& \text{for} & U/W> 1
\end{array}
\right.
\quad .
\label{DcDOS}
\end{equation}

At this point a peculiarity of our model becomes evident:
the charge velocity {\em increases} as a function of $U/W$
and eventually diverges at $U=U_c=W$.
The effective charge mass~$m_c^*$ which is connected
to the density of states or the velocities by
$m_c^*/m = D_c(0)/D_0(0) = v_F/|v_c^L|$
{\em decreases} as a function of $U/W$. This is in contrast
to the Brinkman-Rice scenario for the Mott-Hubbard transition
in that case $m_c^*/m$ {\em increases} and charge excitations tend to
localize close to the transition~\cite{BrinkmanRice,VollhardtReview}.
Indeed, for the Hubbard model with nearest-neighbor hopping
(cosine dispersion) $m_c^*(n,U)$ diverges at the MCIT
for $U>0$ when the transition is approached from below
half-filling ($n\to 1^-$)~\cite{Schulz}.

\subsubsection{Less than half-filling: $n<1$}
In this case there are only right-moving spinons with velocity~$v_s^R$,
eq.~(\ref{vsr}). The spinons are further restricted to $-\pi/2 < {\cal K}
< {\cal K}_{F}/2$ with ${\cal K}_{F}
=\pi \left( 2n-1\right)$.
The lowest-lying charge excitations are now given by {\em holons} alone
while the doublons are always gaped:
\[
\text{charge-exc:\quad}
\left[ \uparrow \downarrow \right] \ldots \left[ \uparrow \downarrow \right]
\circ\biggr|_{\cal K}
\left[ \uparrow \downarrow \right] \ldots
[ \uparrow \biggr|_{{\cal K}_{F}} \downarrow ]
\circ \ldots \circ \; .
\]
Although holons need not come in pairs, they are still
separated by an even multiple of~$\Delta$ in ${\cal K}$-space.
Rescaling again we find for the charge excitation
\begin{equation}
\epsilon_h ({\cal K}) = \epsilon_s ({\cal K}) - \epsilon_s ({\cal K}_F/2)
+ t ({\cal K}_F - 2 {\cal K}) \quad -\pi/2 < {\cal K} < {\cal K}_F/2 \quad .
\label{holondispersion}
\end{equation}
The holons are gapless at ${\cal K} = {\cal K}_F/2$, and the corresponding
velocity, $v_c^L =
\partial \epsilon_h ({\cal K}) / \partial {\cal K} \Bigr|_{{\cal K}_F/2}$,
leads to
\begin{equation}
v_c^L = - v_F \left( 1 + \frac{U}{\sqrt{(W+U)^2 - 4WUn}} \right)\quad .
\label{vhl}
\end{equation}
In turn, this implies that the density of states for charge excitations
at low energies still remains
\begin{equation}
D_c (E \to 0) = \frac{1}{2\pi |v_c^L|} \quad .
\end{equation}

\subsubsection{Ground-state compressibility and magnetic susceptibility}

The results for the ground state energy density~$e_0 (n)$,
eq.~(\ref{groundstate}), allow us to calculate the chemical potential
$\mu = \partial e_0(n)/\partial n$ and
the $T=0$ compressibility $\kappa =  \partial n/ \partial \mu$.
By turning on a small
external magnetic field~${\cal H}_0$ we can also obtain
$e_0(m,n)$ (see Appendix~\ref{appa},
eq.~(\ref{e0m})), and the magnetic susceptibility,
$\chi = \partial m/ \partial {\cal H}_0$, where
the magnetization density,~$m$,
is related to ${\cal H}_0$
by ${\cal H}_0 = \partial e_0(m,n) / \partial m$.
We summarize our results as follows:

(i) $n<1$ or ($n=1$, $U/W <1$):
\begin{mathletters}%
\begin{eqnarray}
\mu &=& \frac{ U - W(1-2n) - \sqrt{(W+U)^2-4WUn}}{4} \\[6pt]
\kappa &=& \frac{1}{\pi |v_c^L|} =
\frac{2}{W\left( 1 + U/\sqrt{(W+U)^2 - 4WUn}\right)}
\label{kappa}\\[3pt]
m &=&  \chi {\cal H}_{0} \\[3pt]
\chi &= &\frac{1}{\pi v_s^R} = \frac{2}{W} \left( 1 + \frac{U}{W} \right)
\quad .
\label{chismaller}
\end{eqnarray}%
\end{mathletters}%
%
\newpage
(ii) $n=1$, $U/W > 1$:
\begin{mathletters}%
\begin{eqnarray}
\mu (n=1^-) &=& \frac{W}{2} \\[3pt]
\mu (n=1^+) &=& U-\frac{W}{2}\\[3pt]
\Delta\mu_c &=& U-W\\
\kappa &=& 0
\label{compressn1}\\
m & = & \chi {\cal H}_0\\
\chi &= & \frac{1}{\pi v_s^R} + \frac{1}{\pi |v_s^L|} = \frac{4U}{W^2}
\quad .
\label{chihalffilling}
\end{eqnarray}%
\end{mathletters}%
The system is an incompressible charge insulator at half-filling and
$U>W$. Note that in the limit $n \to 1^{-}$ the compressibility
stays {\em finite} for $U>W$ (for $U=W$, $\kappa = 4\sqrt{1-n}/W$
for $n\lesssim 1$). This is, nevertheless,
consistent with eq.~(\ref{compressn1}) since, in contrast with
the situation of the usual
Hubbard model~\cite{Shiba,Schulz}, in our case
the function $n(\mu)$ is not differentiable
at $\mu (n=1^-)$ and $\mu (n=1^+)$.

We further note that eqs.~(\ref{kappa}),
(\ref{chismaller}), and (\ref{chihalffilling})
are the generic behavior expected of a
``Luttinger-Liquid''~\cite{Haldane1981},
a point which we exploit in our
discussion of the ground-state correlation functions
(sec.~\ref{CFT}) and the Drude conductivity
(sec.~\ref{g-ology}).

\subsection{1/r-tJ Model}

Since the case of half-filling has already been considered by
Haldane~\cite{haldane2}, here we will be
mainly interested in the less that half-filled situation. From
the effective tJ Hamiltonian~(\ref{tJeff}) it is easy to see that
the hole kinetic
energy favors all particles to be
as close to ${\cal K}=-\pi$ as possible while the
exchange interaction $J_{\cal K} = (J/4) \left[ \pi^2 -
\left( {\cal K} -\Delta/2\right)^2\right]$
tries to distribute the particles symmetrically around
${\cal K} =0$. We thus expect a transition at some critical~$J_c$
in this case.

In general, the ground state can be represented as
\[
\text{ground state:} \quad
\circ \ldots \circ \biggr|_{{\cal K}_{1}}
\left[ \uparrow \downarrow \right] \ldots \left[ \uparrow \downarrow \right]
\biggr|_{{\cal K}_{2}} \circ \ldots \circ
\]
where ${\cal K}_2 = {\cal K}_1 + 2\pi n$ has to be determined from
the minimization of the ground state energy.
We find (see appendix~\ref{appa})
\begin{mathletters}%
\label{e0tJ}%
\begin{eqnarray}
e_0 (n, J\leq J_c) &=& -\frac{Wn(1-n)}{2} - \frac{J\pi^2n^2(3-2n)}{12}
\nonumber \\[6pt]
 &= & - \frac{J_c\pi^2 n}{12}\left[ \left( \frac{J}{J_c} -1 \right) n(3-2n)
 + 3-3n+n^2\right]
\end{eqnarray}
with ${\cal K}_1=-\pi$ and
\begin{eqnarray}
e_0 (n, J\geq J_c) &=& -\frac{W^2n}{2\pi^2J} - \frac{J\pi^2n(3-n^2)}{24}
\nonumber\\[6pt]
 &= & - \frac{J_c^2\pi^2 n}{24J} \left[ \left(3 - n^2\right) \left(
 \frac{J^2}{J_c^2} -1 \right) + 2 \left( 3- 3n +n^2\right) \right]
\end{eqnarray}%
\end{mathletters}%
with $ {\cal K}_1 = -\pi \left(n + 2W/(\pi^2 J)\right) > -\pi$.
Here, $J_c = 2W/(\pi^2(1-n))$ is the critical coupling.
Note that $J_c$ is proportional to
$1/(1-n)$, i.e., it decreases with increasing (hole) doping.
Below we discuss the two cases, $J<J_c$ and $J>J_c$, separately:

\subsubsection{$J<J_c$}
For $J< J_c$ we can make use of the
$1/r$-Hubbard model results
(sec.~\ref{HMsec}).
In particular, $\epsilon_s ({\cal K}) =
J (\left(\pi/2\right)^{2} - {\cal K}^2)/2$
for $-\pi/2 < {\cal K} < {\cal K}_{F}/2$, and the spinon
velocity at ${\cal K} = -\pi/2$ is
\begin{equation}
v_s^R = J\pi/2 \quad.
\label{vsrtJsmaller}
\end{equation}
It is amusing to
note that this is precisely the result for the
Heisenberg-chain with {\em nearest-neighbor}
interactions~\cite{Cloizeux}.
This is consistent with the fact that
the Gutzwiller projected Fermi sea $|\psi_{0}\rangle
= \hat{P}_{D=0} |\text{Fermi-sea}\rangle$\cite{noteGWF},
the ground state
wave function of the $1/r$-tJ model,
is also an excellent trial state for the
Heisenberg chain with nearest neighbor
interaction~\cite{GrosShiba,DVFG}, as well as for the nearest neighbor
supersymmetric tJ model ($J=2t$)~\cite{Yokoyama}.
As in sec.~\ref{HMsec}
the corresponding low energy density of states is given by
\begin{equation}
D_s (E \to 0) = \frac{1}{2\pi v_s^R} \quad \text{for} \quad J< J_c \quad .
\label{DstJsmaller}
\end{equation}

For $0<J<J_c$,
the holon velocity is
calculated as
\begin{equation}
v_c^L = -\left( 2t + J\pi(2n-1)/2\right) = -(\pi/2) \left[ \left(1-n\right)
\left(J_c-J\right) +nJ\right] .
\label{vcltJsmaller}
\end{equation}
Note that the limit $J \to 0$ is peculiar: precisely {\em at} $J=0$
we have a free gas of holons with
$\epsilon_h({\cal K},J=0) = -2t ({\cal K} - {\cal K}_F/2)$
for $-\pi/2<{\cal K}<{\cal K}_F/2$,
but with allowed ${\cal K}$-values spaced by $\Delta/2$.
On the other hand, the limit
$J \to 0$ only gives {\em half} of the excitations since the ${\cal K}$
values are
now spaced by $\Delta$ rather than $\Delta/2$.
This is because, for $J>0$, half of the $J=0$ excitations develop a gap,
corresponding to the energy required to break a spinon pair,
$J( {\cal K}_F) >0$.
Consequently,
the low energy density of states for charge excitations,
\begin{equation}
D_c(E\to 0) = \frac{1}{2 \pi |v_c^L|} \quad \text{for} \quad 0<J<J_c \quad ,
\label{DctJsmaller}
\end{equation}
is only half as big for $J\to 0$ as for $J=0$.

\subsubsection{$J>J_c$}

For $J>J_c$ the lowest-lying spin excitation can be represented as
\[
\text{spin-exc:}\quad
\circ \ldots \circ
\sigma \biggr|_{{\cal K}}
\circ \ldots \circ
\sigma' \biggr|_{{\cal K}'}
\circ \ldots \circ\biggr|_{{\cal K}_{1}}
\left[ \uparrow \downarrow \right] \ldots \left[ \uparrow \downarrow \right]
\circ \circ \biggr|_{{\cal K}_{2}}
\circ \ldots \circ \quad .
\]
In the present case nothing
can be said about the ground state wave function in terms of the
original Fermions, since
the link to the $1/r$-Hubbard model can no longer be made.
Again, the spinons come in pairs but they can now have {\em arbitrary}
separation
in the region outside of ${\cal K}_1 < {\cal K}, {\cal K}' <
{\cal K}_2 $.
A single spinon has the
excitation energy $\epsilon_s ({\cal K}) = t ({\cal K}- {\cal K}_2) +
J ({\cal K}_2)/2 =t ({\cal K}- {\cal K}_1) + J ({\cal K}_1)/2 $ and thus,
a {\em gap} proportional to $J-J_c$
opens in spinon spectrum:
\begin{eqnarray}
\Delta \mu_s &=& 2 \epsilon_s (-\pi)
\nonumber \\
&=& \frac{\pi^2 (1-n)}{4} \left( 1-\frac{J_c}{J} \right) \left[
J(1+n)- J_c (1-n)  \right] .
\label{Deltamus}
\end{eqnarray}
$J=J_c$ then corresponds to the onset of
a metal-to-spin-insulator transition (MSIT).

For $J>J_c$ a charge excitation is represented by
\[
\text{charge-exc.:}\quad
\circ \ldots \circ
[ \uparrow \biggr|_{{\cal K}_1} \downarrow ] \ldots
\left[ \uparrow \downarrow \right]
\circ \biggr|_{{\cal K}}
\left[ \uparrow \downarrow \right] \ldots \left[ \uparrow \downarrow \right]
\biggr|_{{\cal K}_{2}}
\circ \ldots \circ
\]
and, after rescaling~${\cal K}$,
the excitation energy for ${\cal K}_{1}/2 < {\cal K} < {\cal K}_{2}/2$ is
$\epsilon_h ({\cal K}) = t({\cal K}_{2}- 2{\cal K}) + \left(
J(2 {\cal K}) - J({\cal K}_{2}) \right)/2 $.
The corresponding velocities are
\begin{equation}
v_c^R = - v_c^L = J \pi n/2
\label{vcrtJlarger}
\end{equation}
at ${\cal K} = {\cal K}_{1}/2$ and ${\cal K}={\cal K}_{2}/2$,
respectively.
We see that, for low energies,
the gapless excitations are {\em parity-symmetric}
although the underlying
Hamiltonian itself does {\em not} conserve parity.
The low energy density of states for charge excitations,
\begin{equation}
D_c(E\to 0) = \frac{1}{2 \pi v_c^R} + \frac{1}{2 \pi |v_c^L|}
= \frac{2}{J \pi^2 n} \quad
\text{for} \quad J>J_c \quad.
\label{DctJlarger}
\end{equation}
doubles
at the MSIT.

\subsubsection{Ground-state compressibility and magnetic susceptibility}
\label{tJcomp}
The $T=0$ chemical potential,
compressibility, magnetization, and magnetic sus\-cep\-ti\-bi\-li\-ty
can be calculated from
eqs.~(\ref{e0tJ}), (\ref{e0mtJsmaller}), (\ref{e0mtJbigger}),
and~(\ref{criticalH}) of appendix~\ref{appa} .
As in the case of the $1/r$-Hubbard model we
summarize our results ($J_c = 2W/(\pi^2(1-n))$):

(i) $J<J_c$:
\begin{mathletters}%
\begin{eqnarray}
\mu &=& -\frac{W(1-2n)}{2} - \frac{J\pi^2n(1-n)}{2} \nonumber\\[3pt]
& = & - \frac{\pi^2 (1-n)J_c}{4} \left[ 2n \left( \frac{J}{J_c} -1 \right)
+1 \right]
\label{mutJsmaller} \\[6pt]
\kappa &=& \frac{1}{\pi |v_c^L|} = \frac{2}{\pi^2 \left[ \left(1-n\right)
\left(J_c-J\right) +nJ\right]}
\label{kappatJsmaller}\\[3pt]
m &=&  \chi {\cal H}_{0} \\[3pt]
\chi &= &\frac{1}{\pi v_s^R} = \frac{2}{\pi^2 J} \quad .
\label{chitJsmaller}
\end{eqnarray}%
\end{mathletters}%

(ii) $J>J_c$:
\begin{mathletters}%
\begin{eqnarray}
\mu  &=& -\frac{W^2}{2\pi^2 J} - \frac{J\pi^2}{8} (1-n^{2})\nonumber\\[3pt]
 &=& - \frac{\pi^2 J_c^2 (1-n)}{8J} \left[ \left(1+n\right)
 \left( \frac{J^2}{J_c^2} -1 \right) + 2\right]
\label{mutJlarger}
\\[6pt]
\kappa &=& \frac{1}{\pi v_c^R} + \frac{1}{\pi |v_c^L|}
= \frac{4}{J \pi^2 n}\\[6pt]
\Delta\mu_s &=& \frac{\pi^2 (1-n)}{4} \left( 1-\frac{J_c}{J} \right) \left[
J(1+n)- J_c (1-n)  \right]\\[3pt]
m &=& 0\quad \text{for}\quad {\cal H}_0 < {\cal H}_0^c =
 \Delta\mu_s/2
\\[6pt]
\chi &= & 0 \quad .
\label{chibigger}
\end{eqnarray}%
\end{mathletters}%
The system is a spin insulator for $J/J_c>1$. Note that the
compressibility~$\kappa$ doubles at the transition because
another low-energy charge excitation replaces the spin mode which
becomes gaped at $J=J_c$.

\section{Thermodynamical Properties}
\label{sec3}
As noted in Ref.~\onlinecite{prl}, the effective Hamiltonian,~(\ref{heff}),
is equivalent to
a {\em classical} Ashkin-Teller type model in the presence of
an inhomogeneous ``magnetic field''~\cite{AshkinTeller}, with periodic
boundary conditions.
The thermodynamic
properties of our system can then be calculated
by transfer-matrix techniques~\cite{OnsagerBaxter}.
We define the abbreviations
$S_{{\cal K},\sigma} = \exp \left( -\beta h_{{\cal K},\sigma}^{s}
\right)$,
$D_{{\cal K}} = \exp \left ( -\beta h_{{\cal K}}^{d} \right)$,
$E_{{\cal K}} = \exp \left ( -\beta h_{{\cal K}}^{e} \right)$,
and $P_{{\cal K}}= \exp \left( -\beta J_{{\cal K}} \right)$
where  $\beta = 1/k_{B} T$ ($k_B \equiv 1$)
is the inverse temperature.
The transfer matrix between sites ${\cal K}-\Delta$ and ${\cal K}$,
\[
F_{{\cal K}-\Delta,{\cal K}} = \left( \begin{array}{cccc}
S_{{\cal K},\uparrow} & S_{{\cal K},\downarrow}P_{{\cal K}}^{-1}
& D_{{\cal K}} & E_{{\cal K}} \\
S_{{\cal K},\uparrow} & S_{{\cal K},\downarrow}
& D_{{\cal K}} & E_{{\cal K}} \\
S_{{\cal K},\uparrow} & S_{{\cal K},\downarrow}
& D_{{\cal K}} & E_{{\cal K}}P_{{\cal K}} \\
S_{{\cal K},\uparrow} & S_{{\cal K},\downarrow}
& D_{{\cal K}} & E_{{\cal K}} \\
\end{array}
\right) \quad ,
\]
has two vanishing eigenvalues; the remaining
two eigenvalues are given by
\[
{\lambda_{\cal K}}^{\pm} = \frac{1}{2} \left[
X_{{\cal K}} \pm
\sqrt{ X_{{\cal K}}^{2} -4 \left[
 S_{{\cal K},\uparrow}S_{{\cal K},\downarrow}
 \left( 1 - P_{{\cal K}}^{-1} \right)
+  D_{{\cal K}} E_{{\cal K}} \left( 1 - P_{{\cal K}} \right) \right] } \,
\right]
\quad ,
\]
where we introduced the abbreviation
$X_{{\cal K}} = S_{{\cal K},\uparrow} + S_{{\cal K},\downarrow}
+ D_{{\cal K}} + E_{{\cal K}}$.
The partition function for all
chemical potentials~$\mu$,
magnetic fields~${\cal H}_{0}$, interaction strengths~$U/W$, and
temperatures~$T$,
\begin{equation}
{\cal Z}_L = \left( \prod_{-\pi < {\cal K} < \pi} {\lambda_{\cal K}}^{+}
\right) +
\left( \prod_{-\pi < {\cal K} < \pi} {\lambda_{\cal K}}^{-}\right)
\qquad ,
\end{equation}
leads, in the thermodynamic limit
($N_{\sigma}, L \to \infty$, $n_{\sigma}
=N_{\sigma}/L = \text{fixed}$),
to the free energy density
$f(\mu,{\cal H}_{0},U/W,T) =
\lim_{L\to\infty} \left[ -\ln {\cal Z}_L / (\beta L)\right]$,
\begin{equation}
f(\mu,{\cal H}_{0},U/W,T) = - \frac{1}{\beta} \int_{-\pi}^{\pi}
\frac{d {\cal K}}{2\pi} \ln {\lambda_{\cal K}}^{+}
\quad .
\label{freenergy}
\end{equation}
Here we made use of the fact that
${\lambda_{\cal K}}^{-} < {\lambda_{\cal K}}^{+}$ {}~\cite{warning}.
This general form (\ref{freenergy}) remains valid
for the $1/r$-tJ~model~(\ref{tJ})
if we set $D_{\cal K} \equiv 0$. From~eq.~(\ref{tJeff}) it is easy to see that
at half-filling the problem is
equivalent to an Ising model on a ring with
nearest-neighbor coupling,
$J_{\cal K}=(J/4) (\pi^{2}-({\cal K}- \Delta/2)^2)$, and
one thus recovers all of
Haldane's results for the spin-chain~\cite{haldane2}.

\subsection{1/r-Hubbard model at half filling}

First note that at half-filling ($n=1$, i.e.~\cite{howtodo1}, $\mu(T) =U/2$)
without external field the spectrum is completely
specified in terms of {\em independent} spin and charge excitations.
In fact, the free-energy in that case has the simple
spin-charge separated form
\begin{eqnarray}
f (n=1,U/W,T) & = & - U/2 + e_{0}(n=1) \nonumber \\[6pt]
 & & - \frac{2}{2\pi\beta} \int_{-\pi/2}^{\pi/2} d{\cal K}
 \left\{ \ln \left[ 1 + \exp \left(
- \beta \epsilon_{s}({\cal K}) \right) \right] +
\ln \left[ 1 + \exp \left( - \beta \epsilon_{c}({\cal K}) \right) \right]
\right\}
\label{freenergyn=1}
\end{eqnarray}
where the dispersion relations
for the spin (``up" and ``down'' spinons) and charge
(holons and doublons) excitations
were already given in eqs.~(\ref{spinondispersion}),~(\ref{chargeondispersion})
(see figure~\ref{fig1}).
Note that the rescaling of ${\cal K}$-values gives an additional factor
of two in front of the integral such that~(\ref{freenergyn=1})
corresponds to the result of two independent free fermion systems
for charge and spin.

\subsubsection{Specific heat}
The spinon and chargeon densities of states, $D_{s,c} (E)
= 1/(2\pi) \int_{-\pi/2}^{\pi/2}
d{\cal K} \delta\left( E-\epsilon_{s,c}({\cal K})\right)$,
are shown in figure~\ref{fig2}a
and~\ref{fig2}b, respectively.
Already before the MCIT the density of states for the spinons
develops a van-Hove singularity at~$U=W/2$. This
could be interpretated as the formation of the lower (spinon)
and upper (chargeon) Hubbard band which are, however, not yet separated.

Given the densities of states, the internal energy
density $u(n=1,T)$ can be
obtained from $u(n=1,T)
 = e_0(n=1) + 2 \int_{0}^{\infty} d E \left[ D_s(E) + D_c (E)\right]
E f(E)$, where $f(E)=[\exp(\beta E) +1]^{-1}$ is the Fermi-Dirac
distribution function.
The specific heat $c_v = -\beta^{2} \partial u/\partial \beta$ is then
given by
\begin{equation}
c_v(n=1,T) = 4 T \int_{0}^{\infty} dx \left( D_{s} (2xT) +
D_c (2xT)
\right) \left( \frac{x}{\cosh x} \right)^{2} \quad .
\end{equation}
At low temperatures this reduces to
$c_v(n=1,T\to 0) = \gamma(n=1) T$ with
\begin{equation}
\gamma(n=1) =  \frac{\pi^{2}}{3}
\left[ D_s (0) + D_c (0) \right] \quad .
\label{problem}
\end{equation}
This Luttinger Liquid relation~\cite{Haldane1981},
which remains valid for all fillings,
will allow us to identify the conformal charge
of the conformal field theory that governs
the low energy behavior of our model (see sec.~\ref{CFT}).

The behavior of the Sommerfeld coefficient~$\gamma$ as a function
of the interaction parameter, $U/W$, can be extracted from
eqs.~(\ref{spinonvelocities}), (\ref{DsDOS}),
(\ref{vcl}), and (\ref{DcDOS}):
\begin{equation}
\gamma (n=1) = \frac{\pi^{2}}{3} D_0(0)
\left\{ \begin{array}{cll}
1 & \text{for} & U/W < 1 \\
U/W & \text{for} & U/W > 1
\end{array} \right.
\label{gamma}
\end{equation}
where $D_0(0)= 2/W$ is the non-interacting density of states.
Note that ~$\gamma(n=1)$ is {\em unrenormalized}
below the MCIT. The reason is that the {\em increase} in the density
of states for spin excitations is exactly compensated by the
{\em decrease} in the density of states for the charge excitations.
This precise cancellation of the two contributions is
a peculiarity of our model and does not occur in the
conventional Hubbard model.
In a realistic scenario for the Mott-Hubbard transition we expect
a growing effective charge mass (or, equivalently, $D_c(0)$)
because the transport of charge becomes more difficult
due to the Coulomb repulsion~\cite{Schulz}.
Also, the spin transport should become less
effective because the spin exchange energy smoothly reduces from
${\cal O}(t)$ to ${\cal O}(J=4t^2/U)$ and we thus
also expect an increasing
effective spin mass (or $D_s(0)$). Consequently,
there should be an increase of~$\gamma$  below the MCIT
in any realistic Hubbard-type
model.

The specific heat as a function of temperature is shown in fig.~\ref{fig3}.
The general structure of $c_v (T)$ reflects
the behavior of the density of states (see fig.~\ref{fig2}).
Even before the MCIT, at $U=W/2$,
the specific heat develops a two peak structure which reflects the
van-Hove singularity in the density of states for spin excitations.
As already seen in eq.~(\ref{gamma}), the Sommerfeld factor is
continuous at the transition.
Well above the transition ($U>>W$) the lower and upper Hubbard bands
are well separated, which is reflected
in a narrow low-temperature
peak in $c_v (T)$ and a broad maximum around $T={\cal O}(U)$.
In spite of the peculiarities of our model
the overall shape of $c_v(T)$ is expected to be generic for
any Hubbard model with a smooth dispersion relation, in the absence of
perfect nesting.

\subsubsection{Compressibility}
Next we discuss the isothermal
compressibility~$\kappa(T)= \partial n / \partial \mu$ which is
related to the
particle number fluctuation by $\kappa (n=1,T)
= T \langle \bigl( \Delta\hat{N} \bigr)^2 \rangle/L
= - \left( \partial^2/\partial \mu^2\right)f (\mu, T) \biggr|_{\mu=U/2}$.
Here, $\Delta\hat{N} \equiv \hat{N} - \langle
\hat{N}\rangle$.
A direct calculation gives
\begin{equation}
\kappa (n=1,T) = \frac{\beta}{\pi} \int_{-\pi/2}^{\pi/2} d{\cal K}
\left[ \exp \left( -\beta \epsilon_s({\cal K}) \right)
+ \exp \left( \beta \epsilon_c({\cal K}) \right) \right]^{-1}
\quad .
\label{kappaT}
\end{equation}
Note that the compressibility probes the system slightly away from
half-filling.
It is thus seen that not only the chargeon
dispersion $\epsilon_c({\cal K})$ but also the spinon
dispersion~$\epsilon_s({\cal K})$
enters the
expression for the compressibility.
Therefore, strong charge-spin separation, in the sense of a completely
decoupled response of charge and spin excitations to an external
force, does not exist even at half-filling.
It is only for $T \to 0$
that the spinons do not contribute to the
compressibility (or holons to the magnetic susceptibility),
as can be seen from eqs.~(\ref{kappa}), ~(\ref{chismaller}),
and (\ref{chihalffilling}).

The fluctuation of the particle density at half-filling is shown
in figure~\ref{fig4}. At low temperatures and below the MCIT
the fluctuations are linear in temperature, i.e.,
the compressibility is constant and given by~(\ref{kappa}).
Above the transition the charge gap opens, and there are only exponentially
small particle number fluctuations at low temperatures.
At high temperatures the fluctuations saturate at $ T \kappa (n=1, T\to \infty)
= 1/2$. This is the classical value for spin-1/2
electrons on a lattice where on average half of the sites are
doubly occupied or empty.

\subsubsection{Magnetic susceptibility}
The magnetic susceptibility at zero external magnetic field
is given by $\chi (T) =
- \left( \partial^2/\partial {\cal H}_0^2\right)f ({\cal H}_0, T)
\biggr|_{{\cal H}_0=0}$. It can be directly evaluated at half-filling
in terms of both spinon and chargeon degrees of freedom as
\begin{equation}
\chi (n=1,T) = \frac{\beta}{\pi} \int_{-\pi/2}^{\pi/2} d{\cal K}
\left[ \exp \left( \beta \epsilon_s(\kappa) \right)
+ \exp \left( - \beta \epsilon_c({\cal K}) \right) \right]^{-1} \quad .
\label{chiT}
\end{equation}
The result is plotted
in figure~\ref{fig5}.
At low temperatures
it shows Pauli behavior which is strongly enhanced by the interaction,
especially above the MCIT
(see eqs.~(\ref{chismaller}), (\ref{chihalffilling})).
With increasing $U/W$,
due to the enhanced density of spin excitations at low energies,
the susceptibility develops a strong peak at low temperatures.
At high temperatures the susceptibility shows Curie behavior,
$\chi (n=1, T \to \infty) = 1/(2T)$,
the classical value for spin-1/2 electrons
on a lattice where on average half of the sites are singly occupied.
This behavior of $\chi (T)$ is familiar from the Hubbard model
with cosine dispersion at half-filling~\cite{Shiba}, and is thus
a generic feature of Hubbard-type models.

\subsection{1/r-tJ model}
Away from half-filling there is no compact representation of the free
energy in terms of the spinon and chargeon dispersion and
one must use the general form in
eq.~(\ref{freenergy}), with $D_{{\cal K}} \equiv 0$,
and $P_{{\cal K}}= \exp \left( -\beta J \left( \pi^{2} - {\cal K}^{2}\right)/4
\right)$. Furthermore, the chemical potential now depends on both
temperature and density.
We will concentrate on a typical
filling factor, $n=0.75$, which corresponds
to a critical value of the coupling~$J_c/W = 0.81$.

\subsubsection{Chemical potential}
The chemical potential as a function of~$T$ for fixed~$n=0.75$
and various values of $J/J_c$ is shown in figure~\ref{fig6}.
It is seen that~$\mu (J,T)$ depends smoothly on~$T$
for $T/J_c < 1$.
Both above and below the MSIT $\mu$ is only weakly temperature
dependent at low temperatures, a behavior which is especially
pronounced for larger values of~$J/J_c$ where
it remains at its $T=0$ value, eq.~(\ref{mutJlarger}),
for all $T/J_c \leq 0.5$.
Although $\mu(J,T)$ develops stronger $T$-dependence at
low temperatures in
the vicinity of the MSIT,
it remains continuous as a function of~$J$ and shows no anomalies
at $J=J_c$.

\subsubsection{Specific heat}
The specific heat as a function of~$T$ for fixed~$n=0.75$
and various values of $J/J_c$
is shown in figure~\ref{fig7}.
There is just one maximum which gradually becomes broader
and shifts to higher
temperatures with increasing coupling strength~$J$.
More revealing features appear in the temperature dependence of the Sommerfeld
coefficient~$\gamma (T) = c_v (T)/T$, shown in fig.~\ref{fig8},
which is nothing but
$(\partial s/\partial T)$, the temperature derivative of the entropy.
First of all, it is seen that the Luttinger Liquid relation
$\gamma (T=0) = (\pi^2/3) (D_s(0) + D_c(0))$ remains valid
(compare eqs.~(\ref{DstJsmaller}),
(\ref{DctJsmaller}), (\ref{DctJlarger}), and
(\ref{problem})).
For low temperatures and
just above the MSIT,
$\gamma (T)$ shows a prominent peak, reflecting
the large density of states for spin excitations
just above the spin gap. As~$J$ becomes larger, this features
broadens and shifts to higher temperatures and is completely
washed out for~$J/J_c \approx 2$.

\subsubsection{Compressibility}

Fig.~\ref{fig9} shows the isothermal compressibility~$\kappa (n=0.75, T)$
for several values of~$J/J_c$,
(a)~below and (b)~above the MSIT.
For~$J<J_c$ the compressibility decreases with increasing~$J$,
especially through the suppression of the peak at
$T\approx 0.3 J_c$. At the transition,
the compressibility at~$T=0$ doubles because of the appearance of
an extra gapless holon excitation in the spectrum (see sec.~\ref{tJcomp}).
Just below the transition this additional density of states for
charge excitations is already present at low but finite energies
and causes a sharp increase
in the slope of~$\kappa (T)$ for low temperatures such that
the compressibility for~$J/J_c=0.99$ is actually higher than
for~$J/J_c=0.74$ in a temperatures region around~$T/J_c =0.1$.
Above the transition the additional density of states
for gapless charge excitations results in
a new low temperature peak,
which broadens and shifts to higher
temperatures (and is eventually completely suppressed)
as~$J$ is further increased above the MSIT.

\subsubsection{Magnetic susceptibility}

Finally, fig.~\ref{fig10} shows the magnetic susceptibility~$\chi (n=0.75,
T)$ for the $1/r$-tJ model for various values of $J/J_c$ below and above
the MSIT.
The susceptibility can be cast into the form
\begin{equation}
\chi (\mu, T)\!=\! \frac{\beta}{\pi} \int_{-\pi}^{\pi} d {\cal K}
\left\{ \exp \left[ 2\beta \left( t{\cal K}-\mu\right) \right] +
4 \exp \left[ \beta \left( t{\cal K}-\mu\right) \right] +
4 \exp \left[ \beta J \left( \pi^2 -{\cal K}^2\right)/4 \right]
\right\}^{-1/2} \, .
\label{chitJtemperature}
\end{equation}
As expected, for $J<J_c$ we find a finite (Pauli) susceptibility
at $T=0$, eq.~(\ref{chitJsmaller}).
Evaluating the linear term~in~$T$ from eq.~(\ref{chitJtemperature})
one finds $\chi (n,T \to 0) = 2/(J\pi^2) + \left[ 8/(J\pi^2)^{2}\right] T +
{\cal O}(T^2)$, independent of~$n<1$.
Close to the transition, however, the temperature region
over which this expansion is valid shrinks to
zero, and the susceptibility thus
seems to have a negative temperature gradient at low temperatures
close to the MSIT ($J/J_c = 0.99$). Just below the transition
we have a strong increase in the density of states for charge excitations
at low energies which already showed up in the compressibility. Consequently,
the density of states for low-energy spin excitations is considerably
reduced. This results in a low-temperature dip of the magnetic susceptibility
close to the MSIT.

For $J>J_c$ the magnetic susceptibility shows activated behavior
reflecting the opening of the spin gap,
$\Delta\mu_s$ (eq.~(\ref{Deltamus})). It is seen that
the curves for $J/J_c=0.99$ and $J/J_c=1.02$ qualitatively
differ from each other only in a region of very low temperatures ($T \lesssim
0.07 J_c$).

\section{Conformal field theory approach and correlation functions}
\label{CFT}
Since we do not know how to express the original electron operators
in terms of the eigenstates of our effective Hamiltonian,
we cannot directly calculate any correlation functions for
our model. Fortunately, away from any phase transitions, our model
belongs to the class of Luttinger Liquids~\cite{Haldane1981} for
which the low temperature/energy behavior is
dominated by two gapless excitations for charge and spin with
linear spectrum and {\em different}
velocities~$v_s$, $v_c$ (charge-spin separation).
It is then natural to attempt to calculate the low energy
behavior of various correlation functions of our model
by recasting our results
within the framework of conformal field
theory~\cite{Belavin,ConformalFT,frahm,kawakami}
and $g$-ology~\cite{Schulz,Solyom1979,Voit1992,Walter1993}
techniques,
both of which have proved extremely powerful in extracting the physics of
the Luttinger Liquid fixed point.
In this section we focus on the former approach and
leave for the next section the discussion of $g$-ology.

It is well known that $T=0$ can often be viewed as the ``critical point''
in two-dimensional classical or one-dimensional quantum
field theories: correlation functions decay algebraically
instead of exponentially for long times and/or distances, i.e.,
there is no intrinsic length scale.
The
{\em assumption} that, in addition to linearizing the fermionic
spectrum, conformal invariance also holds
at low energies/temperatures
restricts the behavior of
the lowest order
$1/L$-corrections
to the ground state energy density ($E^{L} _0$), and the energies
($E^{L} _{h^{\pm}}$) and momenta ($P^{L} _{h^{\pm}}$) of
low-lying states of system of finite size, $L$. In turn, this
information is sufficient to determine the
long range, long time behavior of correlation functions.
[An equivalent approach which is based on a Landau expansion around the ground
state for Bethe-Ansatz solvable problems gives the same
results~\cite{carmelo}].
In particular, for a one component Fermi gas with a linear spectrum
conformal invariance implies:
\begin{mathletters}%
\label{corrections}
\begin{eqnarray}
E_0^L - L \epsilon_0 &=& - \frac{\pi c v}{6L}
\label{corrections0}\\[6pt]
E_{h^{\pm}}^{L} - E_0^L &=& \frac{2\pi}{L} v \left( N^{+} + N^{-}
+ h^{+} + h^{-}\right)
\label{correctionsexc}\\[6pt]
P_{h^{\pm}}^{L} - P_0^L &=& \frac{2\pi}{L} \left( h^{+} - h^{-}+
N^{+} - N^{-}\right) +P_h^{\infty}
\label{correctionsp}
\quad .
\end{eqnarray}%
\end{mathletters}%
Here, $c$~is the conformal charge, $v$~is the velocity of the
the right- and left-moving elementary
excitations, $h^{\pm}$ are their conformal dimensions,
$P_h^{\infty}$ is the momentum of the sound excitations in the
thermodynamic limit, and $N^{\pm}$ are
integers~\cite{ConformalFT,frahm,kawakami}.

A particularly simple way of determining the conformal
charge from the Sommerfeld coefficient
of the specific heat is due to Affleck~\cite{Affleck}.
For a one-component system with
right- and left-moving elementary excitations
\begin{equation}
\gamma = \frac{\pi}{3} \frac{c}{v} = \frac{\pi^2}{3} c D (0) ,
\label{Affleckcv}
\end{equation}
where $D (0)$ is the  density of states for low energies.

The information of eqs.~(\ref{corrections}) and (\ref{Affleckcv})
determines
the large distance~$x$, long-time~$t$ behavior of correlation functions
of the (primary) fields~$\Phi_{h^{\pm}}(x,t)$ as
follows~\cite{ConformalFT,frahm,kawakami}
\begin{equation}
\langle \Phi_{h^{\pm}}(x,t) \Phi_{h^{\pm}}(0,0) \rangle
= \frac{\exp \left( -iP_{h}^{\infty} x\right)}{\left( x -i vt \right)^{2h^{+}}
\left( x +i vt \right)^{2h^{-}}} \quad .
\label{corrfunctions}
\end{equation}

We are interested in the long-range behavior of the spin-spin
correlation function $C^{\rm SS} (r,t)$, the density-density correlation
function $C^{\rm NN} (r,t)$, and the one-particle Green's function
$G_{\sigma}(x,t)$,
\begin{mathletters}%
\begin{eqnarray}
C^{\rm SS} (r,t) &=& \frac{1}{L} \sum_{l=1}^{L} \langle
\left( \hat{n}_{l+r,\uparrow}(t) - \hat{n}_{l+r,\downarrow}(t) \right)
\left( \hat{n}_{l,\uparrow} - \hat{n}_{l,\downarrow} \right) \rangle
\\[3pt]
C^{\rm NN} (r,t) &=& \left[ \frac{1}{L} \sum_{l=1}^{L} \langle
\left( \hat{n}_{l+r,\uparrow}(t) + \hat{n}_{l+r,\downarrow}(t) \right)
\left( \hat{n}_{l,\uparrow} + \hat{n}_{l,\downarrow} \right) \rangle \right]
- \left( \frac{n}{2}\right)^2
\\[3pt]
G_{\sigma} (r,t) &=& \frac{-i}{2\pi} \frac{1}{L}
\sum_{l=1}^{L} \langle {\cal T} \hat{c}_{l,\sigma}^{\phantom{+}}
\hat{c}_{l+r,\sigma}^{+} (t) \rangle
\end{eqnarray}%
\end{mathletters}%
where ${\cal T}$ is the time-ordering operator.
The only remaining task in computing
the asymptotic behavior of these correlation functions is
identifying the appropriate
(primary) fields associated with the physical operators of interest.

In what follows we use our exact results
to carry out the procedure outlined above.

\subsection{1/r-Hubbard model}

\subsubsection{Conformal Charge}
The obvious generalization of~(\ref{Affleckcv}) to a two-component
system with charge and spin excitations is to
replace $D(0)$ by $(D_s(0)+ D_c(0))$.
As could have been expected,
comparing with eq.~(\ref{problem}) gives~$c=1$~\cite{onecomponent}.

We note that, since
this identification requires finite spin and charge velocities,
it breaks down at the MIT.
This could have been inferred from the excitation spectra for charge and spin
(eqs.~(\ref{spinondispersion}), (\ref{chargeondispersion}))
which are no
longer linear near ${\cal K}= \pi/2$ but behave as $\omega \propto
k^{\alpha}$ with $\alpha =1/2$.
This is reflected in the finite size corrections to the
ground state energy which behave as $\sqrt{1/L}$ instead of
$1/L$. More precisely, from eq.~(\ref{appe0}), one obtains for half-filling
\begin{equation}
(E_0^L -L \epsilon_0) (n=1)  = \sqrt{\frac{1}{L}} \frac{W}{2}
\sum_{s=1}^{\infty} (-1)^s \left( \frac{1}{\pi s} \right)^{3/2}
\int_{0}^{\pi s L} dy \frac{\sin y}{\sqrt{y}} \quad .
\end{equation}
A similar breakdown of conformal invariance occurs in the
context of superintegrable
chiral $N$-state Potts models~\cite{Pottsmodels} for $N\geq 3$.

It is also worth pointing out that, as shown in appendix~\ref{appb},
even away from the transition,
there are some {\em additive} corrections to the
usual formula~(\ref{corrections0}) in our case.
More precisely, we find
\begin{equation}
E_0^L - L \epsilon_0 = - \frac{\pi}{6L} \left[
v_s^R - v_F + |v_c^L| - v_F \right]
\label{surprise}
\end{equation}
(for $n=1$ and $U>W$ one has to read $v_s^L$ instead of $v_c^L$).
It is easy to see that the corrections guarantee
that there are {\em no} $1/L$ corrections present at $U=0$
where $E_0= L\epsilon_0$ is the exact result for all~$L$.
This requirement follows from the fact that
in our case the entire spectrum of
the kinetic energy operator~$\hat{T}$ can be
obtained from $N-1$ independent \hbox{spin-1/2} SU(2) algebras in each sector
with fixed total momentum~$Q$: $\hat{T}_Q^{N} \equiv t(Q-\pi)
+ W \sum_{i=1}^{N-1} \hat{F}_i^z$ \hbox{($N$ even)}.
Thus, in this case the algebraic structure is much simpler
than the usual Virasoro algebra~\cite{Belavin,ConformalFT};
this invalidates the conventional arguments based on conformal invariance
and justifies the absence of finite size correction in the strict $U=0$ limit.

\subsubsection{Conformal dimensions}

According to
eqs.~(\ref{correctionsexc}), (\ref{correctionsp}), in order
to identify the conformal dimensions we need to consider
those lowest-lying states with the quantum numbers of
the physical excitation of interest.
We will consider two- and one-particle excitations which determine the
long-range behavior of the spin-spin and density-density
correlation functions, and one-particle Green's function, respectively.

\paragraph{$n<1$:}
Away from the transition
the lowest-lying two-spinon excitation is obtained from the ground state
by breaking a $\left[ \uparrow \downarrow \right]$-pair at ${\cal K}=
-\pi +\Delta/2$ and ${\cal K}= -\pi +3 \Delta/2$, and placing the electrons
into the $S=1$, $S^z =0$ state at the same ${\cal K}$ points.
The corresponding excitation energy is
\begin{equation}
E_{h_s^{+}}^{L} - E_0^L = J({\cal K}= -\pi +3\Delta/2)
= \frac{2\pi}{L} v_s^R + {\cal O}\left( 1/L^2 \right) \quad ,
\end{equation}
corresponding to $h_s^{+} =1$ (only right-moving spinons).
Furthermore,
\begin{equation}
P_{h_s^{+}}^{L} - P_0^L = \pi + \frac{1}{2} \left[ \left(
-\pi + \Delta/2 \right) \left( -\pi + 3\Delta/2 \right) \right]
= \frac{2\pi}{L} h_s^+
\end{equation}
where we took into account~\cite{prl} that each bound spin-pair contributes
a momentum of~$\pi$, and each unbound spin at ${\cal K}$ adds
${\cal K}/2$. As a result, $P_{h_s^+}^{\infty} =0$.

The charge excitation can be treated analogously and one finds
(again for $n<1$)
\begin{eqnarray}
E_{h_c^{-}}^{L} - E_0^L &=& (-t)({\cal K}_F -3\Delta/2)
+t ({\cal K}_F +\Delta/2) + J({\cal K}_F -\Delta/2) -
J({\cal K}_F + \Delta/2) \nonumber\\[3pt]
&=& \frac{2\pi}{L} |v_c^L| + {\cal O}\left( 1/L^2 \right) \\[6pt]
P_{h_c^{-}}^{L} - P_0^L &=& \frac{1}{2} \left[ - \left(
{\cal K}_F - 3\Delta/2 \right) + \left( {\cal K}_F + \Delta/2 \right)
\right] = \frac{2\pi}{L},
\end{eqnarray}
implying $h_c^- = 1$, $P_{h_c^-}^{\infty} = 0$.

Finally, the one-particle excitation from the $N$-particle ground state
is given by the $N+1$-particle ground state:
\[
\text{ground state ($N+1$ particles):}\quad
\sigma\, [\uparrow \downarrow] \ldots
[\uparrow \biggr|_{{\cal K}_F} \downarrow] \circ \ldots \circ
\]
We then find,
\begin{eqnarray}
E_{N+1,0}^{L} - E_{N,0}^L &=&
t ({\cal K}_F +\Delta/2)
+ \mathop{{\sum}'}_{{\cal K}= -\pi +3\Delta/2}^{{\cal K}-\Delta/2}
\biggl( J({\cal K}) - J({\cal K} + \Delta) \biggr)  \nonumber\\[6pt]
&=& \mu (n) + \frac{2\pi}{L} \left[ \frac{v_s^R}{4} + \frac{|v_c^L|}{4}\right]
 + {\cal O}\left( 1/L^2 \right) \\[9pt]
P_{N+1,0}^{L} - P_{N,0}^L &=& \frac{1}{2} \left[  \left(-\pi
+\Delta/2 \right) + \left( {\cal K}_F + \Delta/2 \right) \right] =
\frac{\pi}{L} + \pi(n-1) \quad ,
\end{eqnarray}
and correspondingly, $h_N^c=h_N^s=1/4$, $P_N^{\infty}
= \pi (n-1) = k_F^e$ ($k_F^e$ is the electronic Fermi momentum).

\paragraph{$n=1$:}
Below the transition ($U/W < 1$) the same results as for $n<1$ apply.
Above the transition ($U/W>1$) we have no gapless charge excitations.
This implies that all correlation function involving charge
excitations (density-density correlation
function, one-particle Green's function) decay exponentially
because their excitation energies always involve the
charge gap~$\Delta\mu_c$.

On the other hand, we have additional spin excitations.
Besides the excitation with two right-moving spinons
($h_s^+ =1$, $P_{h_s^+}^{\infty} = 0$)
we also find the corresponding excitation with two left-moving
spinons ($h_s^- =1$, $P_{h_s^-}^{\infty} = 0$). Furthermore, we
may split the two spinons to form the excited state
\[
\text{spin-exc.:}\quad \sigma [\uparrow \downarrow] \ldots
[\uparrow \downarrow] \sigma'
\]
leading to
\begin{eqnarray}
E_{h_s^{\pm}}^{L} - E_0^L &=&
\mathop{{\sum}'}_{{\cal K}= -\pi +3\Delta/2}^{\pi-\Delta/2} J({\cal K})
- \mathop{{\sum}'}_{{\cal K}= -\pi +5\Delta/2}^{\pi-3\Delta/2} J({\cal K})
 \nonumber\\[6pt]
&=& \frac{2\pi}{L} \left[ \frac{v_s^R}{4} + \frac{|v_c^L|}{4}\right]
 + {\cal O}\left( 1/L^2 \right) \\[9pt]
P_{h_s^{\pm}}^{L} - P_0^L &=& \pi + \frac{1}{2} \left[  \left(-\pi
+\Delta/2 \right) + \left( \pi - \Delta/2 \right) \right] = \pi \quad .
\end{eqnarray}
This gives $h_s^+ = h_s^- =1/4$, and $P_{h_s}^{\infty} = \pi$.

\subsubsection{Correlation functions at large times and distances}
\paragraph{$n<1$ or ($n=1$ and $U/W<1$):}
According to the results in the last subsection and the
general formulae from conformal field theory~(\ref{corrfunctions})
we are now in position to deduce the long-range behavior of correlation
functions:
\begin{mathletters}%
\label{CCG}
\begin{eqnarray}%
C^{\rm SS} (x,t)&\sim &  A \left( \frac{1}{x -v_s^R t}\right)^2 \\[3pt]
C^{\rm NN} (x,t) &\sim &  B \left( \frac{1}{x +|v_c^L|t }\right)^2 \\[3pt]
G_{\sigma} (x,t) &\sim & \frac{e^{-ik_F^e x}}{2\pi}
\frac{1}{\sqrt{\left( x-v_s^R t \right) +i/\Lambda_t}}
\frac{1}{\sqrt{\left( x+|v_c^L| t \right) +i/\Lambda_t}}
\end{eqnarray}%
\end{mathletters}%
where $\Lambda_t$ is a cut-off parameter ($\Lambda_t=\Lambda {\rm sgn} t$).

Note that the two-particle correlation functions are of the
Fermi Liquid form, with renormalized velocities.
On the other hand, the one-particle Green's function
displays Luttinger Liquid behavior
involving square-root singularities rather than the conventional
quasi-particle asymptotic form, $1/\left( x-v_F t\right)$.
Consequently, the Fourier
transformed one-particle Green's function shows {\em no} quasiparticle
peak, i.e., there is no contribution
proportional to $\delta (\omega -v_F k)$ in the
one-particle spectral function.

In fact, the form of $G_{\sigma} (\omega, k)$ is very interesting
but rather complicated. A detailed analysis of its properties
was given recently in refs.~\onlinecite{Meden,Voit}.
Nevertheless, we can already see by dimensional analysis
that there is a step-discontinuity in the momentum distribution
$n_{k,\sigma} =
\langle \hat{c}_{k,\sigma}^+ \hat{c}_{k,\sigma}^{\phantom{+}}\rangle
= (-i) \sum_r \exp (i k r) G(r,t=0_-)$ at $k=k_F^e = \pi (n-1)$.
Our model provides a novel example
of a system which displays a discontinuity in
$n_{k,\sigma}$ in the absence of single-electron like quasi-particle
excitations.
Such systems have been termed
``free Luttinger Liquids''~\cite{Haldane1981}
or ``Gutzwiller Liquids''~\cite{prl}.
This unusual behavior would reflect itself, for example, in the
dependence of the Kondo-temperature on the Kondo coupling
impurity embedded in a ``Gutzwiller Liquid"~\cite{DHLee}.

\paragraph{$n=1$ and $U/W>1$:}
The charge-charge correlation function and the one-particle
Green's function decay exponentially, while the spin-spin correlation
function has the asymptotic behavior
\begin{equation}
C^{\rm SS} (x,t)\sim   A_1 \left( \frac{1}{x -v_s^R t}\right)^2
+ A_2 \left( \frac{1}{x +|v_s^L| t}\right)^2
+  A_3 \frac{\exp \left( i\pi x\right)}{\sqrt{\left(x -v_s^R t\right)
\left( x + |v_s^L| t\right)}} \quad .
\label{CSSaftertransition}
\end{equation}
Due to the mixture of right- and left-moving spinons, an additional
structure (``$2 k_F$-oscillations'') appears in $C^{\rm SS}(x,t)$,
indicating strong antiferromagnetic correlations
beyond the MCIT.

\subsubsection{Equal-time correlation functions}
Conformal field theory only allows us to calculate the
long-range (large~$x$ {\em and}~$t$)
behavior of correlation functions.
On the other hand, the large~$x$ behavior at~$t=0$,
involves contributions from {\em all} frequencies. From eqs.~(\ref{CCG})
it is already clear that the low-frequency
modes lead to
a $1/x^2$ decay for the equal-time spin-spin or density-density correlation
function.
What can be said about the high-frequency contributions?

\paragraph{$n<1$ or $n=1$ and $U/W<1$:}
We know that there is a sharp cut-off wave vector for both
charge- and spin-excitations. In particular, no spin-excitations
are possible above ${\cal K}_F = \pi (2n-1)$.
The highest momentum
spin excitation of the ground state is
\[
\text{maximum $Q$ spin-exc.:}\quad
\sigma [\uparrow \downarrow] \ldots [\uparrow\downarrow] \sigma'
\biggr|_{{\cal K}_F} \circ\ldots\circ
\]
with $Q = \pi + \left[ \left( {\cal K}_F -\Delta/2 \right) +
\left( -\pi +\Delta/2 \right)\right]/2 = n \pi$.
This corresponds to a sharp edge in the correlation function
as a function of momentum~$q$ at ~$q=Q$ (for
$n<1$ or $n=1$ and $U/W<1$ the correlation functions do not diverge at~$q=Q$).
As in ordinary Fermi Liquids, this
reflects itself in long-range oscillations
in real space: $C^{\rm SS} (r,t=0) \sim A_1/r^2 + A_2 \cos (\pi n r)/r^2$.
The same argument applies
to the $t=0$ density-density correlation function which also shows
long-range $2 k_F$-oscillations~\cite{fermibody}.
Finally, the single-particle momentum distribution shows step discontinuities
at both ends of the $U=0$ Fermi ``surface".

\paragraph{$n=1$ and $U/W>1$:} In this regime,
the entire large distance behavior
is already contained in
the large momentum-transfer modes which are also gapless.
More explicitly, for large~$r$ we find
\begin{equation}
C^{\rm SS} (r) \sim A_1 \frac{(-1)^r}{r} + A_2 \frac{1}{r^2}
\end{equation}
which shows strong antiferromagnetic correlations above the
MCIT. The slow oscillating decay of $C^{\rm SS} (r)$ corresponds to
a logarithmic divergence of the spin-spin correlation
function in momentum space near $q=\pi$.

\subsection{1/r-tJ model}
\subsubsection{Conformal Charge}

As expected, eq.~(\ref{problem}) implies ~$c=1$~\cite{onecomponent}.
Note that the formulae~(\ref{corrections})
cannot be applied exactly at the MSIT, $J=J_c$.
At the transition we have {\em both} gapless right-moving
spin-excitations {\em and}
gapless right- and left-moving charge-excitations with {\em finite} velocities
$v_s^R=J_c\pi/2$, $v_c^R=-v_c^L=J_c\pi n/2$.
Nevertheless, the correct form of the $1/L$ correction at $J=J_c$,
$\left[ E_0^L - L \epsilon_0 \right](J=J_c) =
-\left[\pi/(6L)\right] \pi n J_c$, does not contain
a contribution from all these three gapless excitations.
Rather, since a finite density of spin-excitations produces an
effective gap for the charge excitations and vice versa,
only one of the two excitations, spin or charge, should be taken
into account at $J=J_c$.
More precisely, the factor, $\pi n J_c$ in the $1/L$ correction
to the energy can be written in
either of two forms,
$\pi n J_c = \hbox{$v_s^R (J\to J_c^-)$}
+ \hbox{$|v_c^L(J \to  J_c^-)|$} - \hbox{$|v_c^L(J=0)|$} $
(for the origin of the term $v_c^L(J=0)$
see the discussion for the $1/r$-Hubbard model, eq.~(\ref{surprise})),
or $\pi n J_c = v_c^R(J\to J_c^+) + |v_c^L(J\to J_c^+)|$.
In other words, the correct value of the $1/L$ corrections is obtained by
taking the limit from either above or below the transition.

\subsubsection{Conformal dimensions}

\paragraph{$J< J_c$:}
We can again use all results from the $1/r$-Hubbard model in the limit
of large $U/W$. For two-particle excitations
we then obtain $h_s^+=1$, $P_{h_s^+}^{\infty} = 0$;
$h_c^-=1$, $P_{h_c^-}^{\infty} = 0$. Note that for half-filling we
have additional excitations which give $h_s^{\pm}=1/4$,
$P_{h_s^{\pm}}^{\infty} = \pi$.
For the one-particle excitations all excitations involving charges
at gaped for $n=1$, while for $n<1$ we have
$h_N^c=h_N^s=1/4$, $P_N^{\infty}
= \pi (n-1) = k_F^e$.

\paragraph{$J>J_c$:}
Since above the transition we have no gapless spin excitations,
all correlation functions involving spin
excitations (spin-spin correlation
function, one-particle Green's function) decay exponentially.
On the other hand, we have additional charge excitations:
besides the right-moving holon near ${\cal K}_1$
($h_c^+ =1$, $P_{h_c^+}^{\infty} = 0$)
we also find a left-moving holon near ${\cal K}_2$
($h_c^{-} =1$, $P_{h_c^-}^{\infty} = 0$).
Furthermore, we can have a left-moving holon near ${\cal K}_1$
and a right-moving holon near ${\cal K}_2$.
These excitations obey
$h_c^{\pm} =1/2$, $P_{h_c^{\pm}}^{\infty} = \pm ({\cal K}_2 -{\cal K}_1)
= \pm 2\pi n$.

\subsubsection{Correlation functions at large times and distances}
For $J< J_c$ we obtain
the same results as for the Hubbard model
away from the transition,
while $J>J_c$ both the spin-spin correlation function
and the one-particle Green's function decay exponentially.
On the other hand, the charge-charge correlation
function behaves asymptotically like
($v_c=v_c^R=-v_c^L=J\pi n/2$)
\begin{equation}
C^{\rm NN} (x,t)\sim   B_1 \left[ \left( \frac{1}{x -v_c t}\right)^2
+ \left( \frac{1}{x +v_c t}\right)^2\right]
+  B_2 \left[ \frac{\exp \left( i2\pi n x\right)}{\left( x -v_ct\right)}
+ \frac{\exp \left( -i2\pi n x\right)}{\left( x +v_ct\right)} \right]
\quad .
\label{CNNaftertransition}
\end{equation}
Note that since in this case large momentum-transfer excitations are
also gapless,
$C^{\rm NN}(x,t)$ displays ``$4 k_F$-oscillations''~\cite{fermibody}.

\subsubsection{Equal-time correlation functions}
For $J<J_c$ we know that the exact wave function is the Gutzwiller-projected
Fermi-sea, in which case all ground state correlation functions are explicitly
known~\cite{DVFG,MV} for all densities~$n$ and distances~$r$.
In particular~\cite{DVFG},
\begin{equation}
C^{\rm SS} (r\neq 0) = \frac{(-1)^r}{\pi r}
\left[ {\rm Si} \left(\pi r\right) -
{\rm Si} \left(\pi r \left(1-n\right)\right) \right]
\end{equation}
where ${\rm Si}(x) = \int_0^x dt \sin t/t$
is the sine-integral. At half-filling,
the asymptotic behavior~\cite{DVFG},  $C^{\rm SS} (r \gtrsim 5) \simeq
(-1)^r/(2r)$, reflects the strong antiferromagnetic correlations.
The formulae for the density-density correlation function
are rather involved but they show the expected large-distance behavior
\begin{equation}
C^{\rm NN} (r) \sim  \frac{B_1}{r^2} +
\frac{B_2}{r^2} \cos \left(\pi n r\right)+
\frac{B_3}{r^4} \cos \left(2\pi n r\right) \quad .
\end{equation}
The momentum distribution is found to have a jump discontinuity~\cite{MV}
of size $n_{k_{F,+}^e}-n_{k_{F,-}^e} = \sqrt{1-n}$.
It is comforting that all general considerations from conformal
field theory are verified in
a limit where a complete description of the ground state properties
is available.

For $J>J_c$, both
spin-spin correlation function and the one-particle Green's function
decay exponentially. The large momentum transfer processes for
the density-density correlation function are no longer
gaped and the large distance behavior even at $t=0$ can be deduced from
conformal field theory. We find
\begin{equation}
C^{\rm NN} (r) \sim 2B_1 \frac{1}{r^2} + 2B_2 \frac{\cos 2\pi n r}{r}
\quad .
\end{equation}
The momentum transform of the density-density correlation function
shows a logarithmic divergence at $q =2\pi n$.

\section{\lowercase{g}-ology approach and charge transport properties in the
\lowercase{1/r}-Hubbard
model}
\label{g-ology}
At low energies/temperatures
normal electronic systems in one dimension can be described
by a continuum field theory~\cite{Solyom1979}.
The essential idea
is to linearize the electron excitation spectrum near the
two Fermi points, resulting in
left- and right-moving
fermions
(momentum and/or energy transfer cutoffs $\Lambda_B$, $\Lambda$,
must then be introduced to regularize integrals).
These two species of fermions interact via several
scattering channels characterized by coupling
constants~$g_i$, $i=1,\ldots,4$ which can also
be spin-dependent \hbox{(``$g$-ology''} Hamiltonian).
The two large momentum transfer processes are described by
$g_1$, which parametrizes
the scattering process which interchanges right- and left-moving particles
(``backscattering''), and $g_3$, which represents
the scattering of two left-moving particles
into two right-moving particles and vice versa (``Umklapp-scattering'').
The remaining processes, described by $g_2$ and $g_4$,
involve a small momentum
transfer, between a left-moving
and a right-moving electron ($g_2$),
and between electrons on the same branch ($g_4$).
The model with $g_2$ and $g_4$ only is the Luttinger model~\cite{Luttinger}
which has been solved exactly~\cite{MattisLieb,Solyom1979,Haldane1981}.

In general, there is no recipe to link
the $g$-ology coupling constants to the parameters of
a given lattice Hamiltonian
without solving the lattice model exactly~\cite{Schulz}.
Moreover, this identification is only valid for interactions
strengths smaller than the cutoff.
In the case of the Hubbard model with cosine dispersion
the Mott-Hubbard transition happens at half-filling
for $U=0_+$~\cite{LiebWu}, so that the entire low-energy physics,
including the physics of the transition, can be described within
$g$-ology~\cite{Schulz}.
More generally, this approach is applicable to
a wide class of one-dimensional electron systems, the so-called Luttinger
Liquids, the
low energy behavior of which is controlled by a weak coupling fixed
point~\cite{Haldane1981}.
Below we will restrict ourselves to the discussion of the
metallic phase ($n<1$, or $n=1$ and $U/W<1$) of the $1/r$-Hubbard model,
although a similar treatment can be also given for
the metallic phase of the $t-J$ model.

\subsection{Identification of the g-ology parameters}

The $1/r$-Hubbard model is particularly simple as it describes
only right-moving electrons which are, however, hopping {\em on a lattice}.
It is thus immediately evident that, in the context of $g$-ology,
the corresponding low-energy physics
is described by a {\em pure $g_4$-model} (``chiral Luttinger model"),
with a Hamiltonian
\begin{equation}
\hat{H}_{g_4} = \sum_{k,\sigma} (\hbar v_F k)\, \hat{a}_{k,\sigma}^{+}
\hat{a}_{k,\sigma}^{\phantom{+}} + \frac{1}{L} \sum_q \frac{1}{2}
\sum_{\sigma,\sigma'} g_4^{\sigma,\sigma'}\, \hat{\rho}_{\sigma}(q)
\hat{\rho}_{\sigma'}(-q) \quad .
\label{ghamilt}
\end{equation}
Here, $\hat{\rho}_{\sigma}(q) = \sum_k \hat{a}_{k-q,\sigma}^{+}
\hat{a}_{k,\sigma}^{\phantom{+}}$,
the system volume (length of the ring)
is $V=L a$, $\hbar v_F \equiv t a$,
and all terms of the Hamiltonian are understood
to be normal ordered with respect to the ground state
of the non-interacting Hamiltonian,
$\hat{H}_0
= \hat{H}_{g_4\equiv 0}$.
The coupling matrix in (\ref{ghamilt}) can
be decomposed as $g_4^{\sigma,\sigma'} =
g_4^{\parallel}\delta_{\sigma,\sigma'} + g_4^{\perp}\delta_{\sigma,-\sigma'}$.
For sufficiently small values of $U$ ($U/W<<1$)
the lattice plays no role, and
we may obviously
identify $g_4^{\parallel} = {\cal O}(U^2)$, $g_4^{\perp} = U$.

After bosonization~\cite{Haldane1981,Solyom1979,MattisLieb} the Hamiltonian
(\ref{ghamilt}) becomes {\em diagonal}
in the new bosonic operators for charge ($\hat{\alpha}_k$)
and spin ($\hat{\beta}_k$) and thus,
the pure $g_4$-model
is a ``non-interacting'' Luttinger Liquid or
a ``Gutzwiller Liquid"~\cite{prl}.
As discussed above, such a model does have a jump discontinuity
in the momentum distribution although it is not a Fermi Liquid.
The bosonic version of the $g_4$-Hamiltonian now reads
\begin{mathletters}%
\begin{eqnarray}%
\hat{H}_{g_4}^c & = & \sum_k (u_c k)\,
\hat{\alpha}_k^{+} \hat{\alpha}_k^{\phantom{+}}\\
\hat{H}_{g_4}^s & = & \sum_k (u_s k)\,
\hat{\beta}_k^{+} \hat{\beta}_k^{\phantom{+}},
\end{eqnarray}%
with
\begin{eqnarray}%
|u_c| & = & v_F \left[ 1 + \frac{g_4^c}{2\pi\hbar v_F a^{-1}}\right] \\[3pt]
|u_s| & = & v_F \left[ 1 + \frac{g_4^s}{2\pi\hbar v_F a^{-1}}\right] \\[3pt]
g_4^c & = & g_4^{\parallel} + g_4^{\perp}\\
g_4^s & = & g_4^{\parallel} - g_4^{\perp} \quad .
\end{eqnarray}%
\end{mathletters}%
To leading order in $U/W$, $g_4^c =U = -g_4^s$,
and thus, $|u_c| = v_F(1+U/W)$, $|u_s|= v_F
(1-U/W)$~\cite{Voit1992}.
Note that the lattice provides
the natural cut-off parameters, $\Lambda=W$ and $\Lambda_B=\pi/a$.

It is amusing that, for our model, there is a way of extending
the $g$-ology solution beyond the perturbative regime.
In particular, by identifying the velocities, $|u_{c,s}|$, with
the velocities of the charge and spin excitations found in our exact solution,
eqs.~(\ref{vhl}), (\ref{vsr}), we can extract parameters $g_4^{c,s}$
for all values of $U/W$ ($U/W \le 1$ for $n=1$).
The resulting expressions read ($n=N/La$):
\begin{mathletters}%
\label{gident}
\begin{eqnarray}%
g_4^c & = & U \frac{W}{\sqrt{\left(W+U\right)^2 - 4 WUna}} > 0 \\[3pt]
g_4^s & = & - U \frac{W}{U+W} < 0 \quad .
\end{eqnarray}%
\end{mathletters}%
For $n a =1$ (half-filling) $g_4^c \to \infty$ for
$U \to W_-$, reflecting the MCIT in the exact solution.
The identification~(\ref{gident}), for which we give a different argument
in appendix~\ref{appc} (for $n=1$), allows us to follow the solution all the
way to the MCIT by using the $g$-ology parametrization.
This clarifies the fact that, in our model, the MCIT is
not caused by the
the relevance of $g_3$-processes at half-filling
as in the Hubbard model with cosine dispersion (perfect nesting property),
but rather, it arises as a pure renormalization of $g_4^c$ due to the
presence of the lattice.

\subsection{Conductivity at zero temperature: Drude weight}

In this section we make use of the $g$-ology approach to
calculate the zero temperature Drude weight, $D_c$, of the
zero-frequency peak of the real part of the conductivity,
\begin{equation}
{\rm Re} \left[ \sigma (\omega) \right] =
{\cal D}_c \delta (\omega) \quad \text{for}
\quad \omega \to 0.
\end{equation}
In the non-interacting limit,
${\cal D}_{0} = e^2 t a^2/(\pi \hbar^2)$.

To obtain the electrical conductivity we
start with the charge-charge
Green's function
($\hat{\rho}(q) =
\hat{\rho}_{\uparrow} (q) +\hat{\rho}_{\downarrow} (q)$ is the density
operator in momentum space),
\begin{equation}
\chi^{\rm NN} (q,t) = \frac{-i}{La} \langle {\cal T} \hat{\rho} (q,t)
\hat{\rho}(-q,0)\rangle
\end{equation}
the retarded part of which is easily obtained from
the $g$-ology
analysis~\cite{Solyom1979,Penc,Schulz,Walter1993}:
\begin{equation}
\chi^{\rm NN}_{\rm ret} (q,\omega) = \frac{q}{\pi}
\frac{1}{\omega - v_c^Lq + i\eta}.
\end{equation}
This form becomes exact for our model for small values of~$q$ and~$\omega$,
and implies that the low-energy charge transport is
entirely dominated by holons, eq.~(\ref{vcl}).
As a check one may calculate the zero temperature compressibility as
$\kappa (T=0) = \lim_{q\to 0} \lim_{\omega\to 0} \chi^{\rm NN}_{\rm ret}
(q, \omega) = 1/\left( \pi |v_c^L|\right)$ in agreement with
our direct calculation, eq.~(\ref{kappa}).

Up to constant prefactors
which we will put back in the end,
the electrical conductivity follows from the generalized
Einstein relation,
\begin{equation}
\sigma_{\rm ret} (q,\omega) = \frac{i\omega}{q^2} \chi^{\rm NN}_{\rm ret}
(q,\omega),
\end{equation}
which can be derived using the linear response formula for
$\sigma_{\rm ret} (q,\omega)$ and the continuity equation,
$-q \hat{j}(q,\omega) = \omega  \hat{n} (q,\omega)$ (see also
appendix~\ref{appd}).
The real part of the conductivity can then be calculated from
${\rm Re} \left[ \sigma_{\rm ret} (q,\omega) \right]
= |v_c^L| \delta (\omega - v_c^L q)$,
\begin{equation}
\sigma (\omega) = \lim_{q\to 0}
{\rm Re} \left[ \sigma_{\rm ret} (q,\omega) \right] =
|v_c^L| \delta(\omega) \quad.
\end{equation}
Adding back the constant prefactors gives then the final expression for
the Drude weight
\begin{equation}
{\cal D}_c = {\cal D}_{0} \frac {|v_c^L|}{v_F} = {\cal D}_{0} \left[
1 + \frac{U}{\sqrt{\left( W+U\right)^2 - 4WUna}} \right] \quad.
\end{equation}

Note that in our model the Drude weight {\em increases}
with interaction. In fact, at half-filling,
${\cal D}_c$ is seen to {\em diverge} as $\sigma_{\rm dc} (\omega)
= \sigma_0 (\omega) /\left(1-U/W\right)$ for small~$\omega$ and $U/W<1$.
At the same time, the compressibility goes to zero, in such a way that
the product, $\kappa {\cal D}_c$, remains constant.
This behavior is in contrast to the MCIT observed in the Hubbard model
with nearest-neighbor hopping, where, at finite~$U$ the MCIT
can be approached from the metallic state by increasing the filling,
$n\to 1^-$, $U>0$~\cite{Schulz}.
In that case, the compressibility {\em diverges} and the conductivity
tends to zero.

In fact, it is clear that for any realistic dispersion the conductivity
is bounded from above (``f-sum rule''~\cite{fsumrule,ShastrySuth})
and cannot diverge:
$\int_{-\infty}^{\infty} d\omega \, {\rm Re} \sigma (\omega)
\leq F < \infty$.
As shown in appendix~\ref{appd}, $F$ is finite whenever
the bare dispersion
$\epsilon (k)$ has a finite second derivative with respect to~$k$ everywhere.
It is thus seen that it is our discontinuous dispersion relation
$\epsilon (k) = t (k\, \hbox{mod} \, 2\pi)$ which allows for
a diverging conductivity.

Finally, we note that since we do not know how a physical
vector potential couples to the effective particles of our exact solution
one cannot reliably
determine the Drude weight by twisting
the boundary conditions~\cite{Kohn,ShastrySuth}, corresponding to
adding a flux, $\Phi$, through the ring.
In fact,
if we naively
replace all values ${\cal K}$ by ${\cal K} + \Phi/(La)$ to determine
the ground state energy $E_0^L (\Phi)$, and try to evaluate
\begin{equation}
{\cal D} \stackrel{?}{=}
\frac{e^2La}{\hbar^2} \left. \frac{\partial^2 E_{0}^L (\Phi)}{ \partial
\Phi^2}\right|_{\Phi=0}
\label{Drudewrong}
\end{equation}
we would obtain the {\em incorrect} result for the Drude weight,
\begin{eqnarray}
{\cal D} &=& {\cal D}_{0} \left[ \frac{v_s^R}{v_F} -1 + \frac{|v_c^L|}{v_F}
-1 \right] \nonumber\\[3pt]
 &=& {\cal D}_{s} + {\cal D}_c \quad ,
\end{eqnarray}
which involves contributions from both charge {\em and} spin
excitations.

\section{Summary and Conclusion}
\label{summary}
Above we discussed various aspects of the physics of two exactly
solvable models of one dimensional lattice fermions:
the $1/r$-Hubbard model and a related $1/r$-tJ model.
The special feature of these models is that, due to absence of
perfect nesting, they display non-trivial metal-to-charge-insulating
and spin-insulating states, respectively,
at a {\em finite} value of the coupling constants.

The $1/r$-Hubbard model thus provides for an explicit
realization of Mott's and Hubbard's ideas~\cite{Hubbard,Mott}
albeit in a rather pathological model in which the dispersion has
a jump discontinuity at the zone boundaries.
This peculiarity allows for a diverging Drude weight when the
MCIT is approach from the metallic side, associated with
the {\em vanishing} of the charge mass at the transition.
While the behavior in the charge sector in the~$1/r$-Hubbard model
is rather non-generic,
the spin sector reflects the expected physics of correlated Fermi systems:
for example, the magnetic susceptibility strongly increases with interaction
and develops a strong low-temperature peak at the spin-exchange energy
scale~$J=4t^2/U$.
After the MCIT the static spin-spin correlation function displays
strong antiferromagnetic correlations with a logarithmic
divergence in momentum space at~$q=\pi$ .

The ~$1/r$-tJ model with pair-hopping terms
displays a metal-to-spin-insulator transition at~$J_c=
4t/(\pi(1-n))$. The spin liquid state is described by the Gutzwiller projected
Fermi-sea and provides an exactly solvable example
of a genuine RVB-state.
At the MSIT, simultaneously with the opening of a spin gap,
the zero-frequency ``spin-conductivity'' (Drude weight
for spin transport)
drops to zero, after having increased with $J/t$ (to a finite value)
on the spin liquid side.
The MSIT appears
as a pure level-crossing effect, and thus should not be interpreted as
a Mott-type transition~\cite{Mott}.

We also found
that, with a few changes related to the high symmetry of the
spectrum in the non-interacting ($U=0$ and $J=0$) models,
conformal field theory is applicable in the metallic and
spin liquid states (i.e., away from transitions).
Using conformal field theory and $g$-ology techniques
allowed us to calculate long-distance/time properties of
ground state correlation functions. The two-particle sector
shows Fermi Liquid behavior.
However, the single-particle Green's function  does not lead to
a quasi-particle peak in the spectral function, even though
the jump discontinuity in the momentum distribution survives.
Such systems are referred to as
``free Luttinger Liquids'' or ``Gutzwiller Liquids''.
Within the $g$-ology approach the $1/r$-Hubbard model
is identifyed as
a ``pure~$g_4$-model'' or ``chiral Luttinger model''.
In our model the MCIT is driven by short distance lattice effects
which lead to a divergence of
the interaction parameter~$g_4^c$ at the MCIT.

Despite their conceptual shortcomings the two models provide instructive
examples for strongly correlated electron systems. Although a rigorous
proof is still missing we think that our models and results
can be used to check the capability of numerical techniques
and the applicability of various approximations~\cite{GebhardGirndt}.

\acknowledgments
F.G.\ would like to thank his colleagues at Rutgers University for their
hospitality during a visit, and the Deutsche Forschungsgemeinschaft (DFG)
for traveling funds. This research was supported in part
by a grant from the U.S. Department of Energy,
Office of Basic Energy Sciences, Division of Materials Research (D.L.C.),
and by ONR Grant \# N00014-92-J-1378 and a Sloan
Foundation Fellowship (A.E.R.).

\appendix
\section{Calculation of ground state properties in the presence of a magnetic
field}
\label{appa}
In this appendix we calculate some ground state properties in the presence of
a weak magnetic field~${\cal H}_0$, and will eventually let ${\cal H}_0
\to 0$.
\subsection{1/r-Hubbard Model}
In the presence of a magnetic field we will have a finite magnetization.
Spin pairs will be broken and turned
into the direction of the magnetic field.
The positions of these unpaired spins is determined by the spinon
dispersion~(\ref{spinondispersion}). From the discussion above and
figure~\ref{fig1} it is clear that they will
always be located around ${\cal K}=-\pi$. For $n=1$ and $U/W > 1$ there are
also some upturned spins around ${\cal K} =\pi$.
We exemplify the calculation for the latter case.

For $n=1$, $U/W > 1$, and $m>0$ the ground state is represented by
\[
\text{ground state:}\quad
\uparrow\ldots\uparrow\biggr|_{{\cal K}_1}
\left[ \uparrow \downarrow \right] \ldots \left[ \uparrow \downarrow \right]
\biggr|_{{\cal K}_{2}}
\uparrow\ldots\uparrow
\]
where the magnetization~$m$
is given by ${\cal K}_1 - {\cal K}_2 = 2 \pi (m-1)$.
The ground state energy density is calculated as
$e_0 (n=1,{\cal K}_1, {\cal K}_2) =
(1/2\pi) \int_{{\cal K}_1}^{{\cal K}_2} d{\cal K}
\left[ J({\cal K})/2\right]$ where the additional factor $1/2$
took into account that only every second
${\cal K}$ value contributes to the integral.
We are only interested in small fields and expand ${\cal K}_{1,2}
= \mp \pi(1-\eta_{1,2})$ with $\eta_{1,2} \ll 1$.
To second order we then obtain with the help of eqs.~(\ref{groundstate}),
(\ref{vsr}), and~(\ref{vsl})
\begin{equation}
e_0 (\eta_1,\eta_2,n=1) = e_0(n=1) + \frac{\pi}{8}
\left( \eta_1^2 v_s^R + \eta_2^2 |v_s^L| \right) \quad .
\end{equation}
Using $m= (\eta_1 +\eta_2)/2$ we minimize
$e_0 (n=1,\eta_1,\eta_2) -m {\cal H}_0$
with respect to $\eta_{1,2}$, and obtain $\eta_{1,2}
= (2 {\cal H}_0) /(\pi |v_s^{L,R}|)$. The magnetization becomes
\begin{equation}
m= {\cal H}_0 \left( \left(\pi v_s^R)\right)^{-1}
+ \left( \pi |v_s^L|\right)^{-1}\right) = \frac{4 {\cal H}_0 U}{W^2}
\end{equation}
and the ground state energy reads
\begin{equation}
e_{0} (m\to 0,n=1) = e_{0}(n=1) + \frac{1}{8} \frac{W^{2}}{U} m^{2}
\quad .
\end{equation}

A similar calculation gives the ground state energy density
for $n<1$ or $n=1$, $U/W <1$ as
\begin{eqnarray}
e_{0} (m, n<1) &=& \frac{U(n-m) - W\left[ (1-n)n + (1-m)m\right]}{4}
\nonumber \\
&-& \frac{1}{24WU} \left[
\left( \left( W+U \right)^{2} - 4WUm \right)^{3/2}
 - \left( \left( W+U \right)^{2} - 4WUn
\right)^{3/2} \right]
\label{e0m}
\end{eqnarray}
which is valid for all~$m$.
Note that the limits ${\cal H}_0 \to 0$ and $n \to 1$ do not commute
beyond the metal-to-insulator transition,
just as in the usual Hubbard model~\cite{frahm}.

\subsection{1/r-tJ Model}
The ground state representation in the presence of a (not too big)
magnetic field
is given by
\[
\text{ground state:}\quad
\phantom{\biggl|}_{-\pi}\biggr|
\uparrow\ldots\uparrow\biggr|_{{\cal K}_m}
\circ\ldots\circ\biggr|_{{\cal K}_1}
\left[ \uparrow \downarrow \right] \ldots \left[ \uparrow \downarrow \right]
\biggr|_{{\cal K}_{2}}
\circ\ldots\circ\biggr|_{\pi}
\]
with ${\cal K}_2 -{\cal K}_1 = 2\pi(n-m)$ and ${\cal K}_m = \pi (2m-1)$.
A straightforward integration gives $e_0 ({\cal K}_1,{\cal K}_2,{\cal K}_m)
= (-t/(4\pi)) \left[ {\cal K}_1^2 -{\cal K}_m^2 + \pi^2 -{\cal K}_2^2\right]
- (J/(16\pi)) \left[ \pi^2({\cal K}_2-{\cal K}_1) -
({\cal K}_2^3 -{\cal K}_1^3)/3\right]$.
We have to minimize $e_0 ({\cal K}_1,{\cal K}_2,{\cal K}_m)
- \mu n - {\cal H}_0 m$ with respect to ${\cal K}_1,{\cal K}_2,{\cal K}_m$,
where we have to distinguish the cases (i)~${\cal K}_1 = {\cal K}_m$, and
(ii)~${\cal K}_1>{\cal K}_m$ to allow for a transition at some
$J=J_c$.

(i)~${\cal K}_1 = {\cal K}_m$: we are below the transition, and we see that
${\cal K}_2 = \pi (2n-1)$, ${\cal K}_m = \pi (2m-1)$.
In this case we can find
the ground state energy density even without the minimization procedure.
For small external fields it has the simple form
\begin{equation}
e_0(n,m,J<J_c) = - \frac{Wn(1-n)}{2} - \frac{J\pi^2(n-m)}{12}
\left[ 3(n+m) -2 (n^2+nm+m^2)\right] \quad .
\label{e0mtJsmaller}
\end{equation}
By differentiating this expression with respect to~$m$ gives the
connection between the
external magnetic field~${\cal H}_0$ and the (small)
magnetization density~$m$
\begin{equation}
m = {\cal H}_0 \left(\pi v_s^R\right)^{-1} \quad .
\end{equation}

(ii)~${\cal K}_1>{\cal K}_m$: from the minimization equations
one easily finds that ${\cal K}_1 + {\cal K}_2 = - 4W/(\pi J)$.
We thus get a solution
\begin{equation}
{\cal K}_1>{\cal K}_m \quad , \quad
\text{if} \quad J > J_c(m) = \frac{2W}{\pi^2(1-n-m)}
\quad .
\end{equation}
Note that the presence of a magnetic field {\em stabilizes}
the spin-liquid phase where spin-excitations are gapless:
$J_c(m) > J_c(m=0)\equiv J_c$.
The ground state energy density then reads
\begin{equation}
e_0(n,m,J>J_c(m)) = \frac{W^2}{2\pi^2J} (n-m) -\frac{W}{2} m(1-m)
-\frac{J\pi^2}{24} (n-m)
\left[ 3- (n-m)^2\right] \quad .
\label{e0mtJbigger}
\end{equation}
By differentiating this expression with respect to~$m$ gives the
connection between the
external magnetic field~${\cal H}_0$ and the (small)
magnetization density~$m$
\begin{mathletters}%
\label{criticalH}
\begin{eqnarray}
{\cal H}_0 &=& {\cal H}_0^c +
m \pi^2\left[ J_c + n (J-J_c) \right]
\\[3pt]
{\cal H}_0^c &=& \frac{\pi^2(1-n)}{8} \left( 1- \frac{J_c}{J} \right)
\left[ J (1+n) - J_c(1-n) \right] \quad .
\end{eqnarray}
\end{mathletters}%
It now takes a {\em finite} magnetic field~${\cal H}_0^c$ to magnetize
the system. The spin gap is~$\Delta\mu_s=2 {\cal H}_0^c$
(see~(\ref{Deltamus}))
because we break a pair of spins ($S=0$) to form an $S^z=S=1$ state.

\section{1/L corrections for the ground state energy of the
1/\lowercase{r}-Hubbard model}
\label{appb}
The ground state energy for finite system sizes~$L$ (even) and a finite
particle number~$N$ (even) is calculated for $n\leq 1$ as
$E_0^L = - \mathop{{\sum}'}_{-\pi < {\cal K} < {\cal K}_F} J_{\cal K}
+ \sum_{{\cal K}_F < {\cal K} < \pi} (-t) {\cal K}$. The prime
indicates that only every second of the ${\cal K} = \Delta (m + 1/2)$
$m=-L/2,\ldots\ ,L/2 -1)$
has to be taken (${\cal K}_F = \Delta (N-L/2)$).
After rescaling $ m = 2r -L/2 +1$ the sums give
\begin{equation}
E_0^L = L \left[ \frac{Un}{4} - \frac{Wn(1-n)}{4}\right]
- \frac{1}{2} \sum_{r=0}^{N/2-1} \sqrt{W^2 + U^2 - 4WU(2r+1)/L}\quad .
\end{equation}
To calculate the sum we use the Poisson sum formula (see, e.g.,
Ref.~\onlinecite{LLV})
\begin{equation}
\sum_{r=0}^{N/2-1} h(r+1/2) = \int_{0}^{N/2} dx\, h(x)
+ 2 \sum_{s=1}^{\infty} (-1)^{s} \int_{0}^{N/2} dx\,
h(x) \cos \left( 2\pi x s\right) \quad ,
\end{equation}
do the first integral and a partial integration in the second,
and arrive at
\begin{equation}
E_0^L - L\epsilon_0 = - \sum_{s=1}^{\infty} (-1)^s \frac{2WU}{L\pi s}
\int_{0}^{N/2} dx \frac{\sin \left(2\pi x s\right)}{\sqrt{\left( W+U\right)^2
- 8 WUx/L}}
\quad .
\label{appe0}
\end{equation}
Away from $U=W$ and $n=1$ we can do further and further partial integrations
in the integral on the right hand side of this equation, and generate
an expansion in $1/L$. Keeping only the first term gives
\begin{equation}
E_0^L - L\epsilon_0 = - \sum_{s=1}^{\infty} \frac{(-1)^s}{s^2}
\frac{WU}{L\pi^2 (W+U)}
\left[ 1 - \frac{1}{\sqrt{1- 4WUn/\left( W+U\right)^2}} \right]
+{\cal O} \left(1/L^{2}\right)
\quad .
\end{equation}
Doing the remaining sum and rearranging terms ($v_F =t= W/(2\pi)$
is the Fermi velocity)
gives
\begin{equation}
E_0^L - L\epsilon_0 = \frac{\pi}{6L} v_F
\left[ \frac{U}{U+W} - \frac{U}{\sqrt{(W+U)^2 - 4 WUn}} \right] \quad .
\end{equation}
With the help of eqs.~(\ref{spinonvelocities}) and (\ref{vcl})
this can be cast into the form of eq.~(\ref{surprise}).

\section{Effective \lowercase{g$_4$}-parameters at half-filling from
the screened electron-electron interaction}
\label{appc}
There is a rather simple recipe to obtain $g_4^{c,s}$ for
all $ U/W<1$ without referring to the exact solution.
At half-filling we know that $k_F^e=0$ lies symmetrically
around the upper and lower band edge ($\pm W/2$ at $k=\pm \pi/a$).
For a pure $g_4$-model one can exactly calculate
the {\em screened} electron-electron interaction
with the help of Ward-identities~\cite{Solyom1979,Penc,Walter1993}.
One finds
\begin{equation}
D_4^{c,s} (k,\omega) = g_4^{c,s} \frac{\omega -v_F k}{\omega - |u_{c,s}|k}
\quad .
\end{equation}
The effective interaction integrated over all times is given by the
$\omega =0$ contribution which gives
\begin{equation}
D_4^{c,s} (k,0) = g_4^{c,s} \frac{v_F }{|u_{c,s}|} = g_4^{c,s}
\frac{1}{1 + g_4^{c,s}/W}
\quad .
\end{equation}
If we now {\em demand} that the
time-integrated part of the screened interaction
has to stay unrenormalized at its value for small~$U/W$,
$D_4^c(k,0) = U$ and $D_4^s(k,0) = -U$, we find that
\begin{mathletters}%
\begin{eqnarray}%
g_4^c = U \frac{1}{1-U/W} \\[3pt]
g_4^s = - U \frac{1}{1+U/W}
\end{eqnarray}%
\end{mathletters}%
which is indeed the correct result. So far, we have not been able to
obtain a general form for $g_4^{c,s}$ away from half-filling,
without appealing to the exact solution.

\section{\lowercase{f}-sum rule for general dispersion relations}
\label{appd}
\subsection{Current operator}
For a Hamiltonian $\hat{H}= \hat{T} + \hat{V}$ we assume
that the interaction only contains density operators~$\hat{n}_r$.
We omit spin indices in the following.
The particle density operator
$\hat{n}(q) = \sum_{r} e^{iqr}\hat{n}_r = \sum_{k}
\hat{c}_{k+q}^+\hat{c}_{k}^{\phantom{+}}$ then commutes with~$\hat{V}$.
We expand the exponential in the particle density operator
$\hat{n}(q)$ for small momenta~$q$ (this step could cause problems
in the case of long-range hopping). The continuity equation then reads
\begin {equation}
i\frac{\partial}{\partial t}\hat{n}(q) =
-q\sum_{r}\frac{\partial r}{\partial t}
\hat{n}_r = -q\sum\limits_{r} v(r) \hat n(r) \quad .
\label{continuity}
\end{equation}
The sum on the right-hand side is identified with the particle current operator
\begin{equation}
\hat{j}=\lim_{q \to 0} \frac{-i}{q} \frac{\partial}{\partial t} \hat{n}(q)
\quad .
\end{equation}
Using the Heisenberg equations of motion we can calculate $\hat{j}$
as
\begin{equation}
\hat j=\lim_{q \to 0} \sum_{k} \hat c_{k+q}^+ \hat c_{k}
\left( \frac{\epsilon(k+q)-\epsilon(k)}{q} \right)
\quad .
\label{eqmotion}
\end{equation}
If the dispersion relation is differentiable, this reduces to
the standard expression for the current operator involving
the group velocity ${\partial\epsilon(k)}/{\partial k}$.
Our linear dispersion relation
is not differentiable at $|k| =\pi$. Instead we may think that
the linear dispersion relation
is the result of a limiting process
where we start with a differentiable dispersion $\widetilde{\epsilon} (k)$
which becomes ever steeper near $|k| = \pi$.
This corresponds to ever faster
left-moving electrons near $|k|=\pi$,
in addition to the right-moving
electrons for $|k|< \pi$. From this viewpoint
it becomes clear that we have to expect a {\em singular} contribution from
these ``left-moving'' electrons. Indeed, we find for $q>0$
\begin{equation}
\hat{j}(q)=\frac{W}{2\pi} \sum_k \hat{c}_{k+q}^+ \hat{c}_k^{\phantom{+}} -
\frac {W}{q}\sum_{\pi-q<k<\pi} \hat{c}_{k+q}^+ \hat{c}_k^{\phantom{+}}
\end{equation}
and $ \hat j (-q)=\hat j^+ (q)$. Now let $q\to  0$ and assume
that  we may approximately replace
$\hat{c}_{k+q}^+ \approx \hat{c}_k^+$. For the minimum $q=\Delta =2\pi/L$
one has $\hat{j} (q=\Delta)=
W/(2\pi) \left[ \hat{N} - L \hat{c}_{-\pi+\Delta/2}^+ \hat{c}_{\pi-
\Delta/2}^{\phantom{+}} \right]$.
As ``the current at $q=0$'' one may {\em define}
\begin{eqnarray}
\hat{j} &=& \frac {1}{2} \left( \hat{j}(q=\Delta) +
\hat{j} (q=-\Delta)\right) \nonumber \\
&=& W/(2\pi) \left[ \hat{N} - (L/2) \left( \hat{c}^{+}_{-\pi+\Delta/2}
\hat{c}_{\pi-\Delta/2}^{\phantom{+}}
+ \hat{c}^{+}_{\pi-\Delta/2} \hat{c}_{-\pi + \Delta/2}^{\phantom{+}}
\right) \right] \quad .
\end{eqnarray}
We obviously get an extensive (i.e., singular) contribution from the states
near the Brillouin zone boundary.

\subsection{f-sum rule}
We are interested in the real part of the transverse conductivity for small
$q$ at temperature $T=0$
\begin{equation}
{\rm Re} \left[ \sigma (\omega, q \to 0) \right]
= \frac {e^2}{L\omega} \lim_{q \to 0}
{\rm Re} \left\{ \int_{-\infty}^{t} dt'e^{i \omega (t-t')}
\langle \left[ \hat{j}^{+} (q,t) , \hat{j} (q,t') \right] \rangle
\right\} \quad .
\end{equation}
Using the continuity equation~(\ref{continuity}),
the equation of motion (in $t$) (eq.~(\ref{eqmotion})),
and integrating by parts (in $t'$) one finds
\begin{eqnarray}
{\rm Re} \left[ \sigma (\omega, q \to 0) \right]
&=& - \frac {e^2}{L \omega q^{2}}
{\rm Re} \biggl\{ (-i) \left. \langle \left[ \left[ \hat{\rho}^{+} (q,t),
\hat{T} \right], \hat{\rho} (q,t') \right] \rangle  e^{i\omega (t-t')}
\right|_{- \infty}^{t}  \nonumber \\
 & &  - (-i)(-i\omega) \int_{-\infty}^{t} dt' e^{i \omega (t-t')}
\langle \left[ \left[ \hat{\rho}^{+} (q,t),
\hat{T} \right], \hat{\rho} (q,t') \right] \rangle \biggr\}\; .
\label{monster}
\end{eqnarray}
The first term in $\{ \ldots\}$ is purely imaginary and drops out.
To obtain the f-sum rule we integrate over all $\omega$.
This gives
\begin{eqnarray}
F (q) &\equiv& \int_{-\infty}^{\infty} d\omega {\rm Re} \left[ \sigma (\omega,
q \to 0) \right] \nonumber \\
 &=&  - \frac {\pi e^2}{L q^{2}}
\langle \left[ \left[ \hat{\rho}^{+} (q),
\hat{T} \right], \hat{\rho} (q) \right] \rangle
\end{eqnarray}
because the expression~(\ref{monster}) only depended on $(t-t')$.
The double commutator gives
\[
\left[ \left[ \hat{\rho}^{+} (q),\hat{T} \right], \hat{\rho} (q) \right]
= \sum_k \hat{c}_k^{+} \hat{c}_k^{\phantom{+}}
\biggl[ \epsilon (k) - \epsilon (k+q)+
\epsilon(k) - \epsilon (k-q)\biggr]
\]
Thus,
\begin{equation}
F(q) = \frac{\pi e^2}{L} \sum_k \left\{ \frac{1}{q}
\left[ \frac {\epsilon (k+q)-
\epsilon (k)}{q} + \frac{\epsilon (k-q) -\epsilon (k)}{q}\right] \right\}
\langle \hat{n}_k \rangle
\end{equation}
and $F = \lim_{q \to 0} F(q)$.
For twice differentiable dispersion relations this reduces to
$F= (\pi e^2/L) \sum_k \langle \hat{n}_k \rangle
\left[ \partial^2 \epsilon (k)\bigr/\partial k^2\right]$. In particular, for a
cosine dispersion, $F= {\pi e^2} \langle - \hat{T} \rangle/{L}$
which is the standard result~\cite{fsumrule,ShastrySuth}.

For the linear dispersion we get
\begin{equation}
F(q)=\frac{\pi e^2 W}{Lq^2} \left\{\sum_{-\pi<k<-\pi+q}
\langle \hat{n}_k\rangle
-\sum_{\pi-q<k<\pi} \langle\hat{n}_k \rangle \right\} \quad .
\end{equation}
The minimum value of $q$ is $\Delta=2\pi/L$.
Thus
\begin{equation}
F \equiv F(\Delta)=
L\frac{e^2 }{4\pi} \Bigl[\langle \hat n_{\pi-\Delta/2}\rangle
-\langle\hat n_{-\pi+\Delta/2} \rangle\Bigr] \quad .
\end{equation}
The f-sum rule is thus {\em extensive} and leaves room for a
diverging Drude weight. This can only happen, if the dispersion relation
$\epsilon (k)$
is not twice differentiable with respect to~$k$.

\begin{figure}
\caption{Dispersion relations for (a) spinons,
see eq.~(\protect\ref{spinondispersion}),
and (b) chargeons, see eq.~(\protect\ref{chargeondispersion}),
in the $1/r$-Hubbard model for (a) $U/W = 0.25, 0.5, 0.75, 1, 2$, and
(b) $U/W = 0.5, 1, 2$.}
\label{fig1}
\end{figure}

\begin{figure}
\caption{Total density of states (a)~$D_{s}$ of the spinons
and (b)~$D_{c}$ of the chargeons
for the $U/W$-values of fig.~\protect\ref{fig1} in the $1/r$-Hubbard model
(note the change in energy scale);
the energy gap in $D_c$ for $U>W$ is only half as big as
the true gap because the chargeons always come in {\em pairs}.}
\label{fig2}
\end{figure}

\begin{figure}
\caption{Specific heat as a function of temperature for the
half-filled $1/r$-Hubbard model
for various values of $U/W$
(a) below ($U/W=0, 0.5, 0.75, 1$),
(b) above ($U/W=1, 2, 4$)
the metal-to-charge-insulator transition.}
\label{fig3}
\end{figure}

\begin{figure}
\caption{Fluctuation of the particle density~$k_{B}T \kappa (n=1,T)$
for the half-filled $1/r$-Hubbard model
as a function of temperature for $U/W=0, 0.5 ,1 ,2$.}
\label{fig4}
\end{figure}

\begin{figure}
\caption{Magnetic susceptibility~$\chi (n=1, T)$
for the half-filled $1/r$-Hubbard model
as a function of temperature for $U/W=0, 0.5 ,1 ,2$.}
\label{fig5}
\end{figure}

\begin{figure}
\caption{Chemical potential~$\mu (n=0.75, T)$ in the $1/r$-tJ~model
as a function of temperature for $J/J_c=0.5, 0.99, 1.02, 1.2 ,2$;
the metal-to-spin-insulator transition happens at~$J_c=0.81 W$.}
\label{fig6}
\end{figure}

\begin{figure}
\caption{Specific heat~$c_v (n=0.75, T)$
in the $1/r$-tJ model
as a function of temperature for $J/J_c=0.49, 0.99 ,1.02 ,2.47$.}
\label{fig7}
\end{figure}

\begin{figure}
\caption{Sommerfeld coefficient~$\gamma (n=0.75, T)$
in the $1/r$-tJ model
as a function of temperature for $J/J_c=0.49, 0.99 ,1.02 ,2.47$.}
\label{fig8}
\end{figure}

\begin{figure}
\caption{Compressibility~$\kappa (n=0.75, T)$
in the $1/r$-tJ model
as a function of temperature (a) below
($J/J_c=0.49, 0.74, 0.99$) and (b) above ($J/J_c=1.02 ,1.23 , 1.73, 2.47$)
the MSIT.}
\label{fig9}
\end{figure}

\begin{figure}
\caption{Magnetic susceptibility~$\chi (n=0.75, T)$
in the $1/r$-tJ model as a function of temperature for
$J/J_c=0.49, 0.74 ,0.99 ,1.02, 1.23, 2.47$.}
\label{fig10}
\end{figure}


\begin{references}
\bibitem{prl} F.~Gebhard and A.E.~Ruckenstein, Phys.~Rev.~Lett.~{\bf 68},
244 (1992).
%
\bibitem{Hubbard} J.~Hubbard, Proc.~R.~Soc. London, Ser.~A~{\bf 276},
238 (1963).
%
\bibitem{fermibody} The diameter of the Fermi body is $\pi n$ as usual.
This is the wave vector
that enters in the long distance behavior of correlation functions, and
is identified with ``$\pi n= 2 k_F$''.
%
\bibitem{HarrisLange} P.W.~Anderson in {\sl Solid State Physics},
ed.\ by F.~Seitz and D.~Turnbull, vol. {\bf 14}, p. 99 (1963);
J.E.~Schrieffer and P.A.~Wolff, Phys.~Rev.~{\bf 149}, 491 (1966);
A.B.~Harris and R.V.~Lange, Phys.~Rev.~{\bf 157}, 295
(1967).
%
\bibitem{fdmh-bss} F.D.M.~Haldane, Phys.~Rev.~Lett.~{\bf 60},
635 (1992); B.S.~Shastry, Phys.~Rev.~Lett.~{\bf 60},
639 (1992).
%
\bibitem{Mott} N.F.~Mott, Rev.~Mod.~Phys.~{\bf 40}, 677 (1968);
{\sl Metal-Insulator Transitions}, (Taylor \& Francis, London, 1974).
%
\bibitem{AndZou} P.W.~Anderson, G.~Baskaran, Z.~Zou, and T.~Hsu,
Phys.~Rev.~Lett.~{\bf 58}, 2790 (1987);
F.C.~Zhang and T.M.~Rice, Phys.~Rev.~{\bf B~37}, 3759 (1988);
F.C.~Zhang, C.~Gros, T.M.~Rice, and H.~Shiba, Supercond. Sci.
and Techn.~{\bf 1}, 36 (1988).
%
\bibitem{singuletts} The configurations $[\bullet \circ]$ and
$[ \uparrow \downarrow]$ represent charge ($C=C^z=0$)
and spin singlets ($S=S^z=0$).
%
\bibitem{WangColeman} D.F.~Wang, Q.F.~Zhang, and P.~Coleman,
(unpublished) (1993).
%
\bibitem{AndersonRVB} P.W.~Anderson, Science {\bf 235}, 1196 (1987).
%
\bibitem{howtodo1} In a grand canonical ensemble
we use $\hat{c}_{k,\sigma}^{+} \mapsto
\hat{c}_{-k,-\sigma}^{\phantom{+}}$. The kinetic energy
operator is invariant, and $\hat{D} - \sum_{\sigma}
\mu_{\sigma} \hat{N}_{\sigma} \mapsto
\hat{D} - \sum_{\sigma} \mu_{\sigma} \hat{N}_{\sigma}
+ (L- \sum_{\sigma} \hat{N}_{\sigma}) (U- \sum_{\sigma}
\mu_{\sigma})$. At half-filling
this must be an operator identity so that we have to fix
$\sum_{\sigma} \mu_{\sigma} = 2\mu = U$.
%
\bibitem{BrinkmanRice} W.F.~Brinkman and T.M.~Rice, Phys.~Rev.~{\bf B~2},
4302 (1970).
%
\bibitem{VollhardtReview} D.~Vollhardt, Rev.~Mod.~Phys.~{\bf 56},
99 (1984).
%
\bibitem{Schulz} H.J.~Schulz, Phys.~Rev.~Lett.~{\bf 64}, 2831 (1990);
Int.~J.~Mod.~Phys.~{\bf B~5}, 57 (1991);
{\sl Proceedings of the Adriatico Research Conference and
Miniworkshop on ``Strongly Cor\-re\-lated Electron Systems~II},
ed. by G.~Baskaran et al. (Progress in High Temperature
Superconductivity -- Vol.~29, World Scientific, Singapore, 1991);
C.F.~Coll, Phys.~Rev.~{\bf B~9}, 2150 (1974).
%
\bibitem{Shiba} H.~Shiba and P.A.~Pincus, Phys.~Rev.~{\bf B~5}, 1966 (1972);
H.~Shiba, Prog.~Theor.~Phys.~{\bf 48}, 2171 (1972);
H.~Shiba, Phys.~Rev.~{\bf B~6}, 930 (1972).
%
\bibitem{Haldane1981} F.D.M.~Haldane, J.~Phys.~C {\bf 14}, 2585 (1981);
Phys.~Rev.~Lett. {\bf 45}, 1358 (1980); {\bf 47}, 1840 (1981).
%
\bibitem{haldane2}
F.D.M.~Haldane, Phys.~Rev.~Lett.~{\bf 66}, 1529 (1991).
%
\bibitem{Cloizeux} J.~des Cloizeux and J.J.~Pearson, Phys.~Rev.~{\bf 128},
2131 (1962).
%
\bibitem{noteGWF} The unique ground state at $U=0$ (Fermi sea)
and for $U >0$ contain one and the
same configuration with no double occupancy.
It is the only one to survive for $U \to \infty$,
hence the Gutzwiller projected Fermi sea is the exact ground state
for the $1/r$-tJ model at all fillings.
%
\bibitem{GrosShiba} H.~Yokoyama and H.~Shiba, J.~Phys.~Jpn.~{\bf 56},
1490 (1987); {\sl ibid.} 3570 (1987); {\sl ibid.} 3582 (1987);
{\sl ibid.} {\bf 57}, 2482 (1988);
C.~Gros, R.~Joynt, and T.M.~Rice, Phys.~Rev.~{\bf B~36}, 381 (1987);
C.~Gros, Phys.~Rev.~{\bf B~38}, 931 (1988).
%
\bibitem{DVFG} F.~Gebhard and D.~Vollhardt, Phys.~Rev.~Lett.~{\bf 59},
1472 (1987); Phys.~Rev.~{\bf B 38}, 6911 (1988).
%
\bibitem{Yokoyama} H.~Yokoyama and M.~Ogata,
Phys.~Rev.~Lett.~{\bf 67}, 3610 (1991).
%
\bibitem{AshkinTeller} J.~Ashkin and E.~Teller, Phys.~Rev.~{\bf 64},
178 (1943).
%
\bibitem{OnsagerBaxter} L.~Onsager, Phys.~Rev.~{\bf 65}, 117 (1944);
see also R.J.~Baxter, {\sl Exactly solved models in statistical mechanics},
(Academic Press, New York, 1982).
%
\bibitem{warning} There is always a true gap between $\lambda_{\cal K}^-$
and $\lambda_{\cal K}^+$ so that the contribution of the
$\lambda_{\cal K}^-$ to the free energy density
does not show up in a $1/L$-expansion.
%
\bibitem{Belavin}
A.~A.~Belavin, A.M.~Polyakov, and A.B.~Zamolodchikov,
Nucl.~Phys.~{\bf B~241}, 333 (1984).
%
\bibitem{ConformalFT}
J.L.~Cardy, Nucl.~Phys.~{\bf B 270} [FS16], 186 (1986);
H.W.~Bl\"{o}te, J.L.~Cardy, and M.P.~Nightingale,
Phys.~Rev.~Lett.~{\bf 56}, 742 (1986);
A.D.~Mirnov and A.V.~Zabrodin, Phys.~Rev.~Lett.~{\bf 66},
534 (1991).
%
\bibitem{frahm} H.~Frahm and V.E.~Korepin, Phys.~Rev.~{\bf B~42},
10553 (1990); {\bf B~43}, 5653 (1991).
%
\bibitem{kawakami}
N.~Kawakami and S.-K.~Yang, Phys.~Lett.~{\bf 148A}, 359 (1990);
Phys.~Rev.~Lett.~{\bf 65}, 2309 (1990);
Prog.~Theor.~Phys.~Suppl.~{\bf 107}, 59 (1992).
%
\bibitem{Solyom1979} J.~S\'{o}lyom, Adv.~Phys.~{\bf 28}, 201 (1979).
%
\bibitem{Voit1992}J.~Voit, Phys.~Rev.~{\bf B~45}, 4027 (1992).
%
\bibitem{Walter1993} W.~Metzner and C.~di~Castro, Phys.~Rev.~{\bf B 47},
16107 (1993).
%
\bibitem{carmelo} J.~Carmelo and A.A.~Ovchinnikov, J.~Phys.~{\bf C~3},
757 (1991); J.M.P.~Carmelo, P.~Horsch, and A.A.~Ovchinnikov,
Phys.~Rev.~{\bf B~45}, 7899 (1992);
J.M.P.~Carmelo and P.~Horsch, Phys.~Rev.~Lett.~{\bf 68},
871 (1992).
%
\bibitem{Affleck} I.~Affleck, Phys.~Rev.~Lett.~{\bf 56}, 746 (1986).
%
\bibitem{onecomponent} Note that our model is chiral, i.e.,
the density of states is only half as big as in the usual case.
%
\bibitem{Pottsmodels} G.~Albertini, B.M.~McCoy, J.H.H.~Perk,
S.~Tang, Nucl.~Phys.~{\bf B~314}, 741 (1989);
R.J.~Baxter, J.~Stat.~Phys.~{\bf 57}, p.~1 (1989).
%
\bibitem{Meden} V.~Meden and K.~Sch\"{o}nhammer, Phys.~Rev.~{\bf B~46}, 15753
(1992), and unpublished (1993).
%
\bibitem{Voit} J.~Voit, Phys.~Rev.~{\bf B~47}, 6740 (1993), and
unpublished (1993).
%
\bibitem{DHLee} D.-H.~Lee and J.~Toner, Phys.~Rev.~Lett.~{\bf 69},
3378 (1992), and private communication.
%
\bibitem{MV} W.~Metzner and D.~Vollhardt, Phys.~Rev.~Lett.~{\bf 59},
121 (1987); Phys.~Rev.~{\bf B 37}, 7382 (1988).
%
\bibitem{Luttinger} J.M.~Luttinger, J.~Math.~Phys.~{\bf 4}, 1154 (1963).
%
\bibitem{MattisLieb} D.C.~Mattis and E.H.~Lieb, J.~Math.~Phys.~{\bf 6}, 304
(1965).
%
\bibitem{LiebWu} E.H.~Lieb and F.Y.~Wu, Phys.~Rev.~Lett.~{\bf 20}, 1445 (1968).
%
\bibitem{Penc} E.H.~Rezayi, J.~Sak, and J.~S\'{o}lyom, Phys.~Rev.~{\bf B 20},
1129 (1979); {\sl ibid.} {\bf B~23}, 1342 (1981);
K.~Penc and J.~S\'{o}lyom, Phys.~Rev.~{\bf B 44}, 12690 (1991).
%
\bibitem{fsumrule} See, e.g., D.~Pines and P.~Nozi\`{e}res,
{\sl Theory of Quantum Liquids} (Benjamin, New York, 1966);
P.F.~Maldague, Phys.~Rev.~{\bf B~16}, 2437 (1977).
%
\bibitem{ShastrySuth} B.S.~Shastry and B.~Sutherland, Phys.~Rev.~Lett.~{\bf
65}, 243 (1990).
%
\bibitem{Kohn} W.~Kohn, Phys.~Rev.~{\bf 133}, A~171 (1964);
for recent applications, see C.A.~Stafford, A.J.~Millis, and B.S.~Shastry,
Phys.~Rev.~{\bf B 43}, 13660 (1991);
R.M.~Fye, M.J.~Martins, D.J.~Scalapino, J.~Wagner, and W.~Hanke,
Phys.~Rev.~{\bf B~45}, 7311 (1992).
%
\bibitem{GebhardGirndt} F.~Gebhard and A.~Girndt, ``Comparison
of Variational Approaches for the Exactly Solvable $1/r$-Hubbard model'',
available as preprint cond-mat/9307036 from the preprint library
{\tt cond-mat\verb2@2babbage.sissa.it}~(accepted for
publication in Z.~Phys.~B).
%
\bibitem{LLV} L.D.~Landau and E.M.~Lifshits, {\sl Statistical Physics, part~1},
 Course on Theoretical Physics Vol.~V, \S~60
(3rd edition, Pergamon Press, New York, 1980).
%
\end{references}
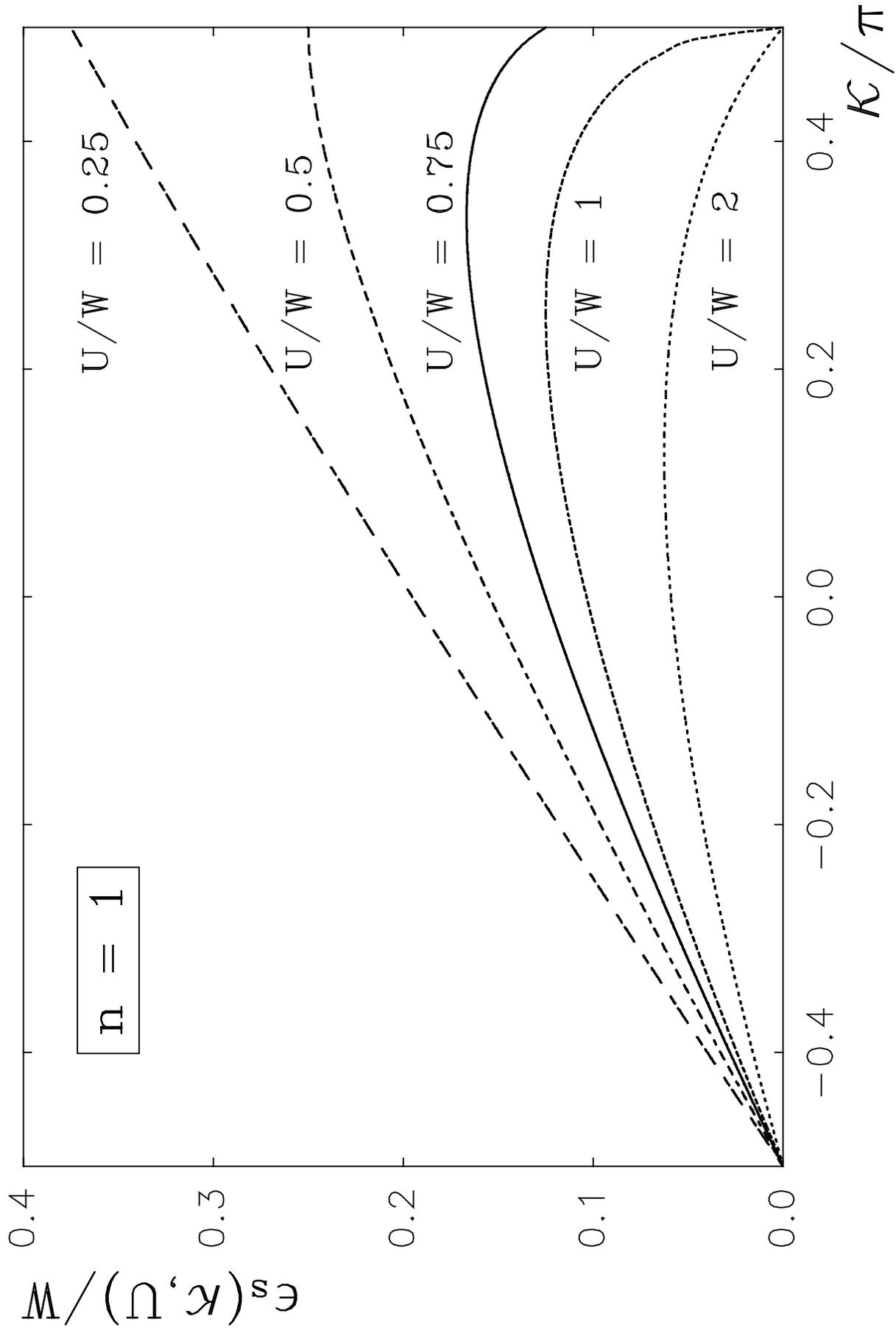
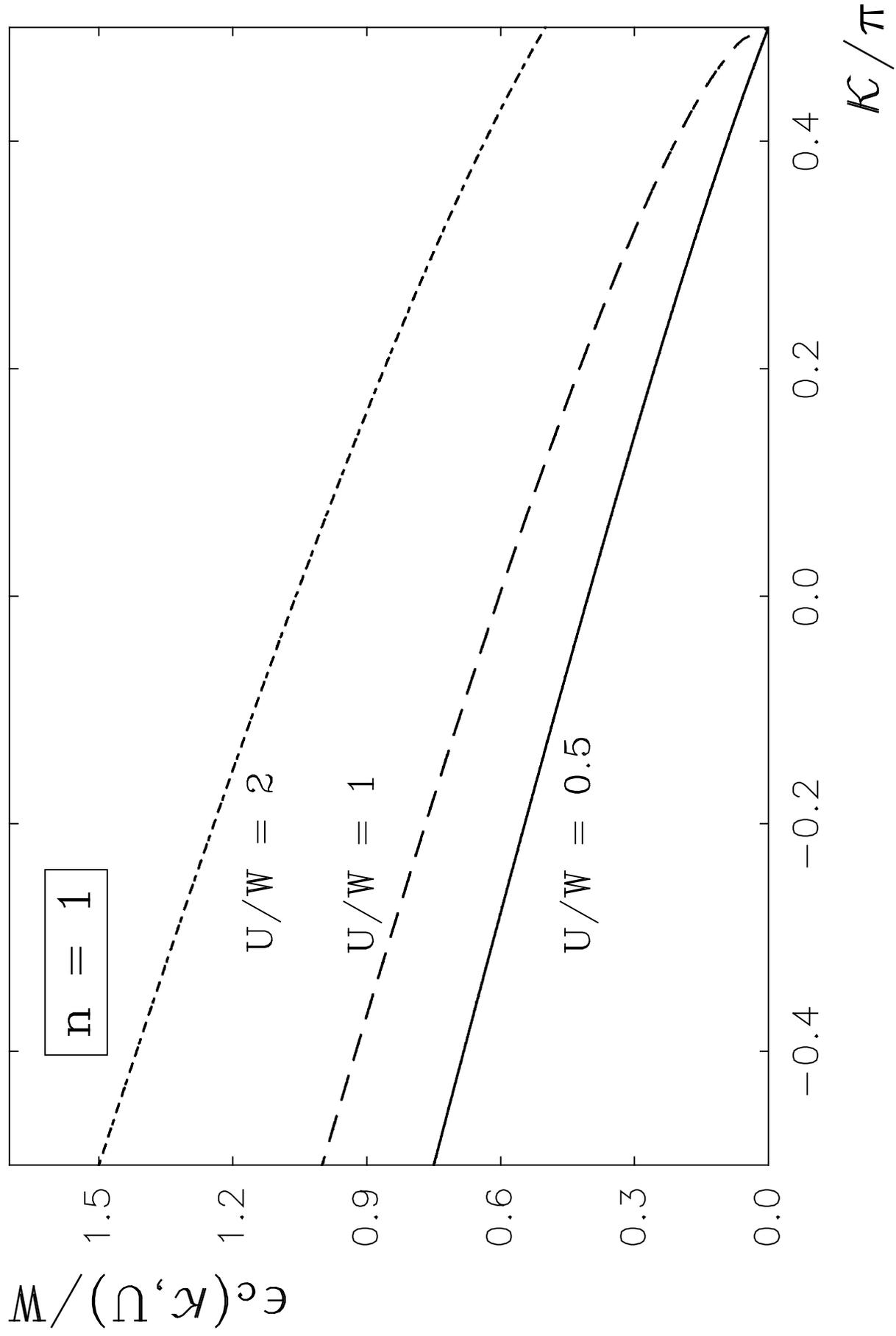
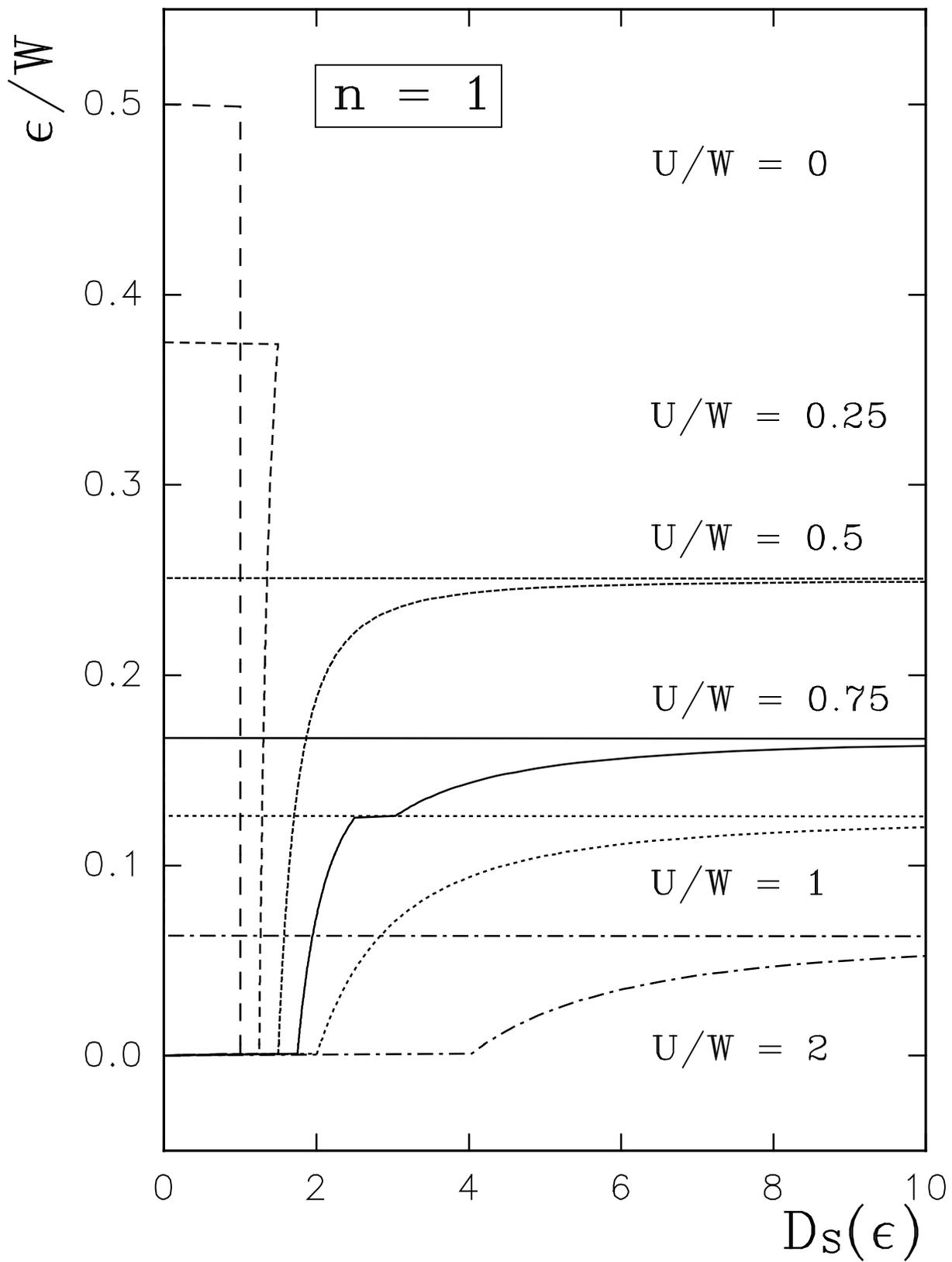
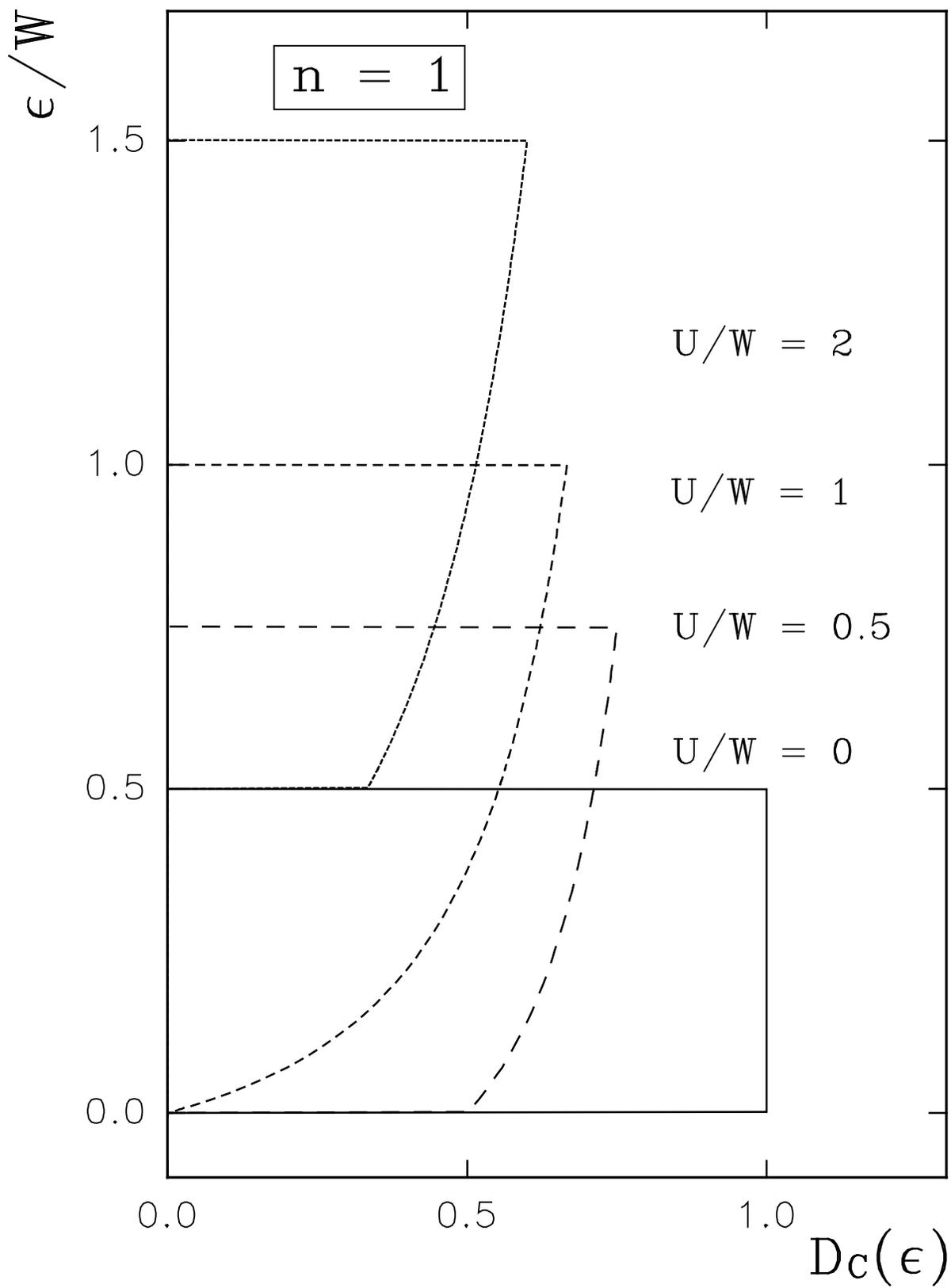
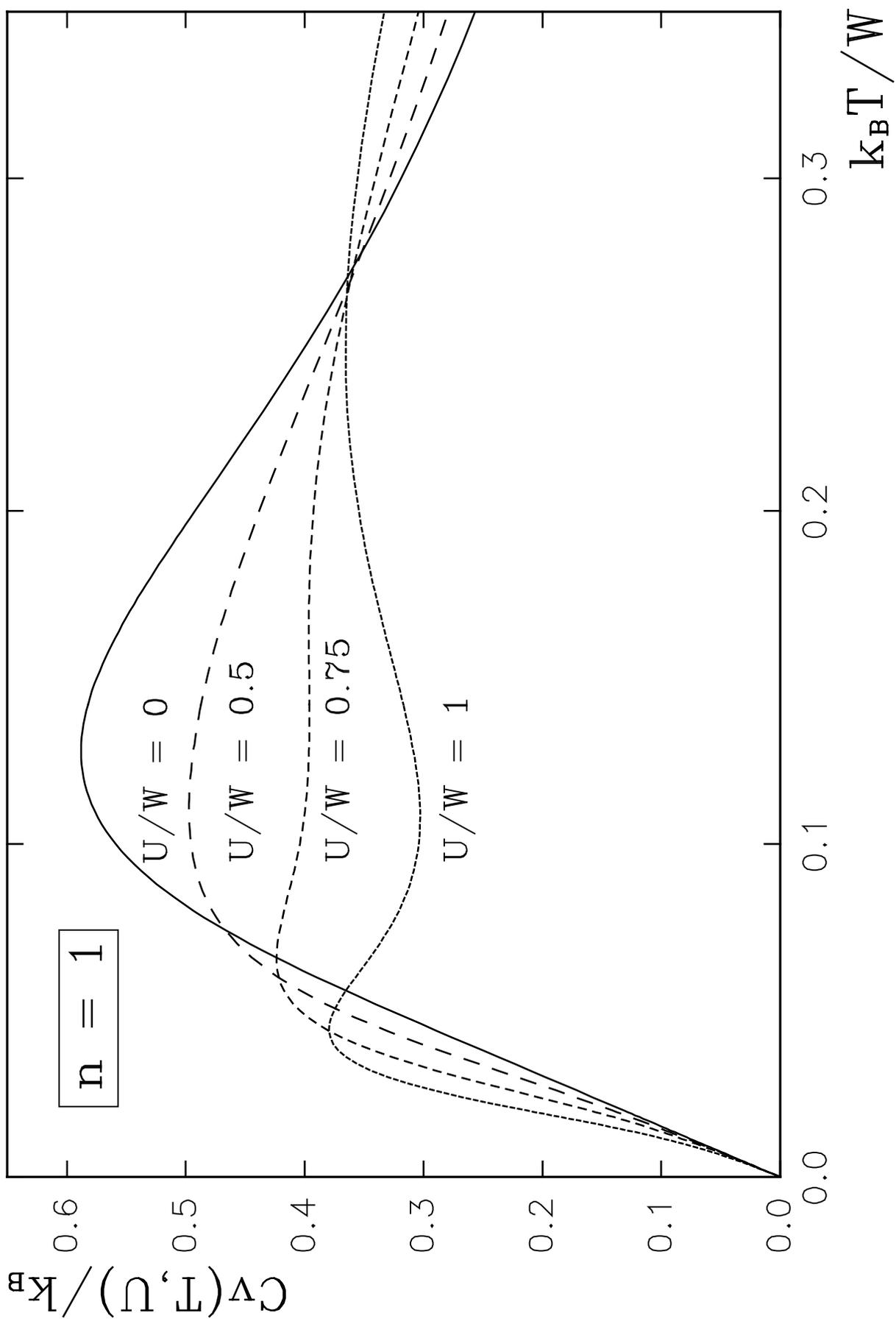
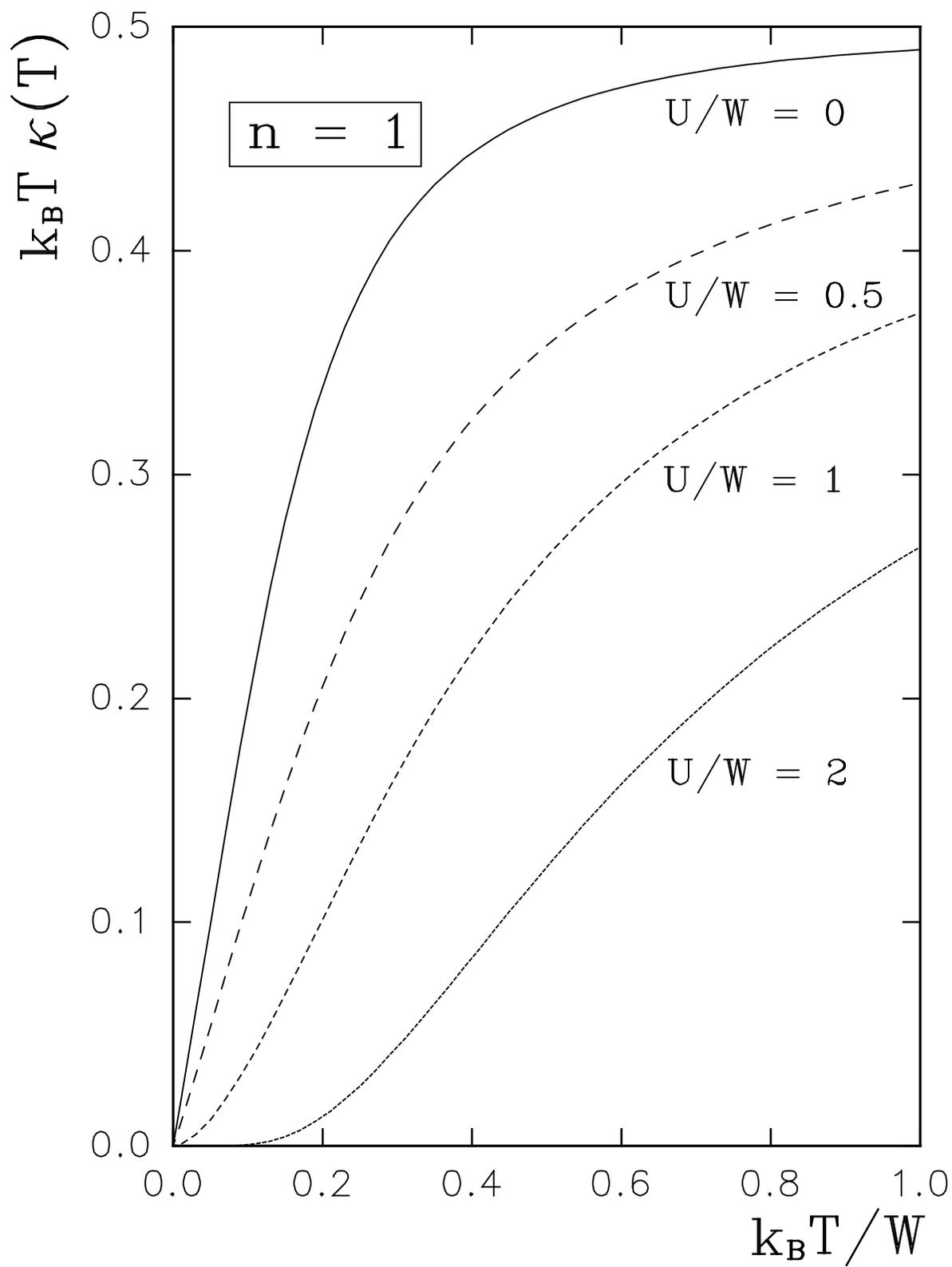
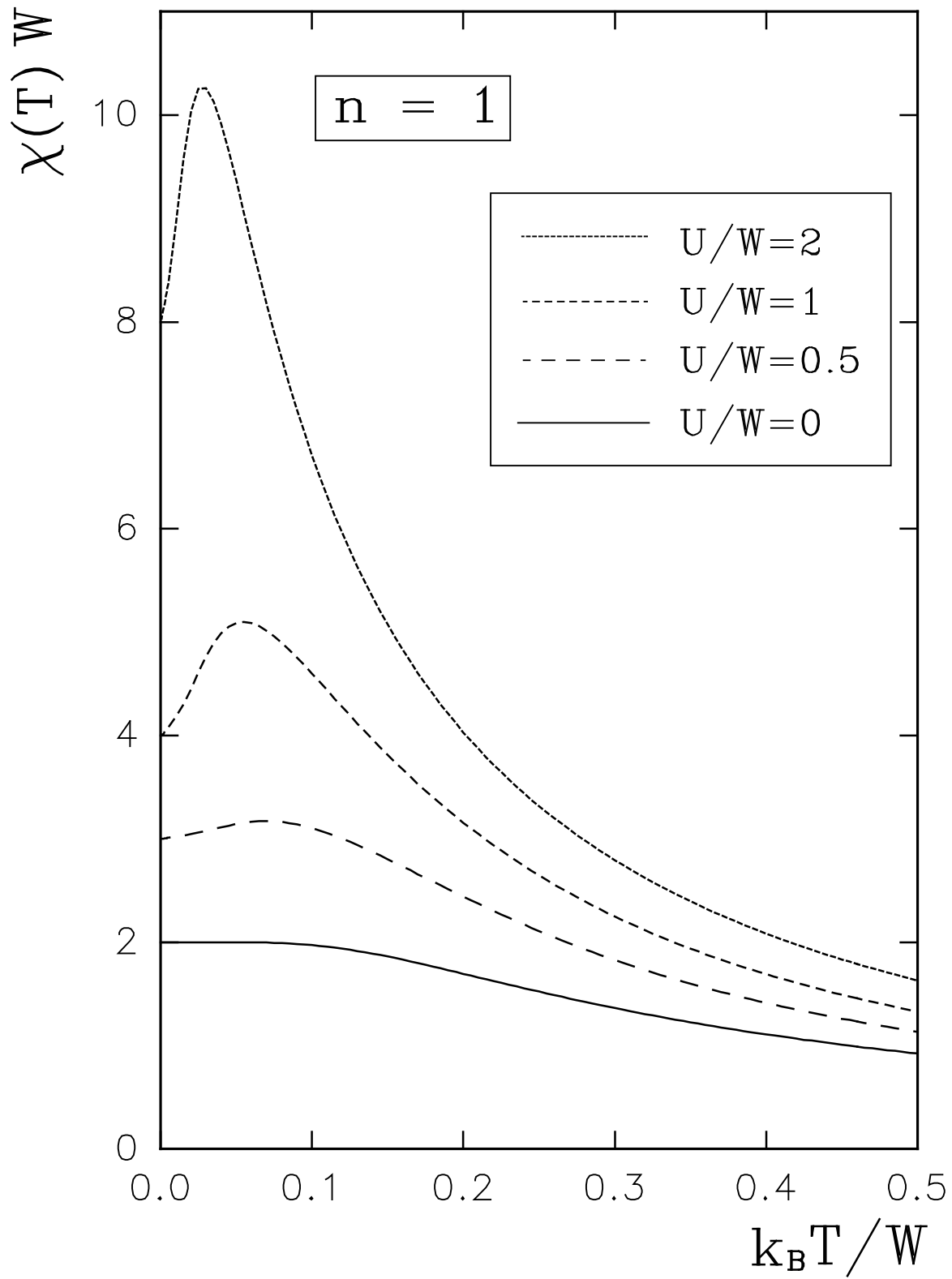
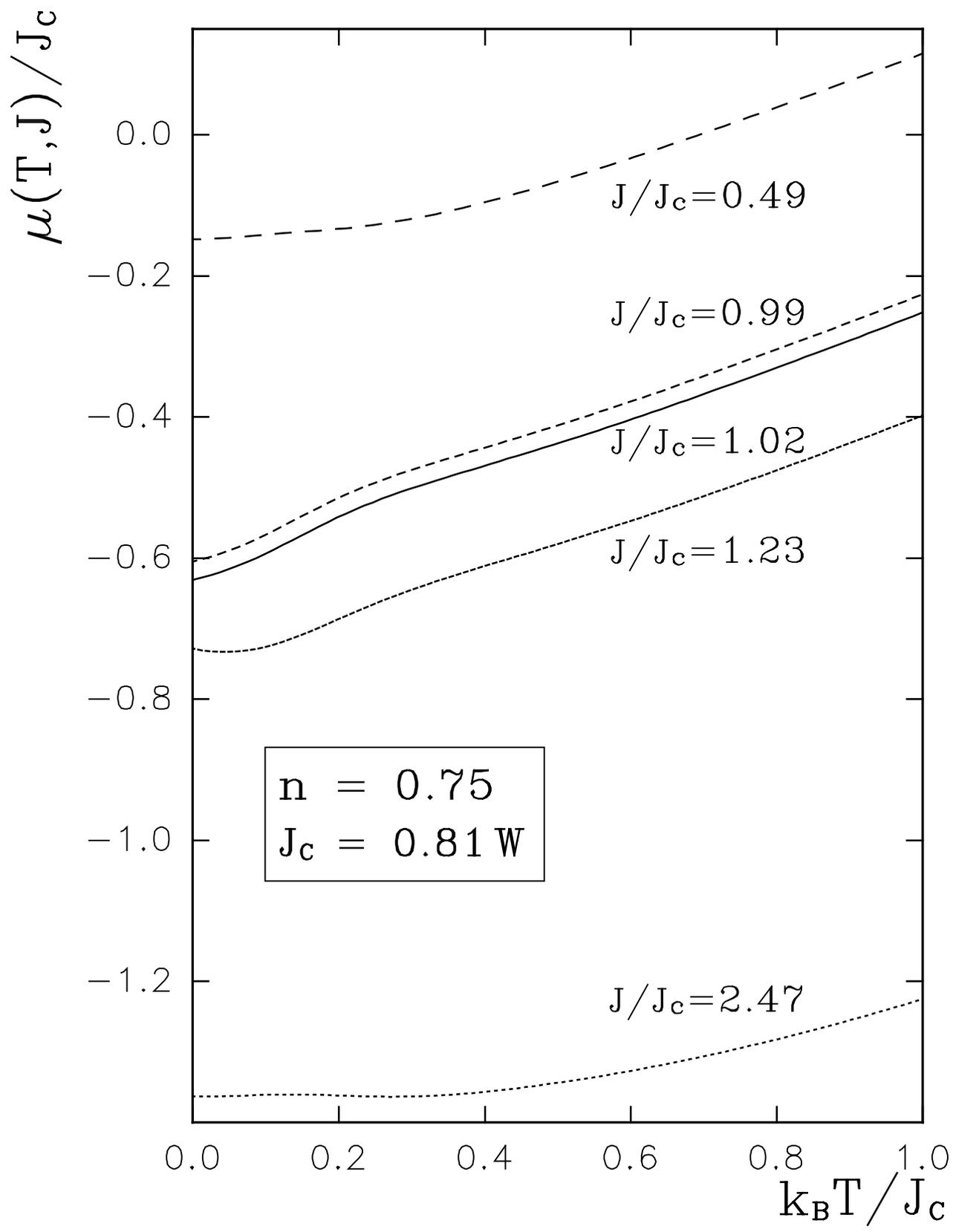
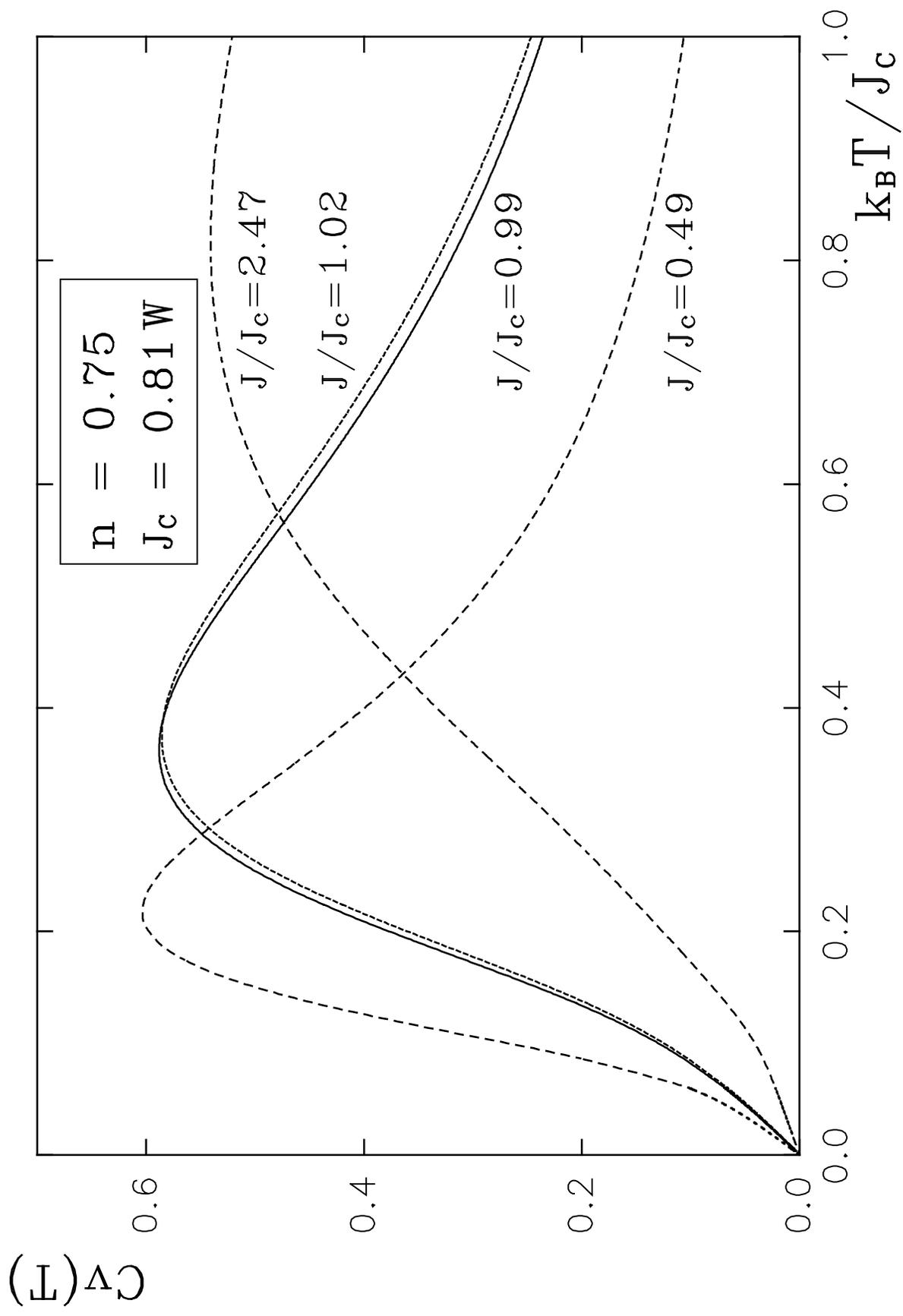
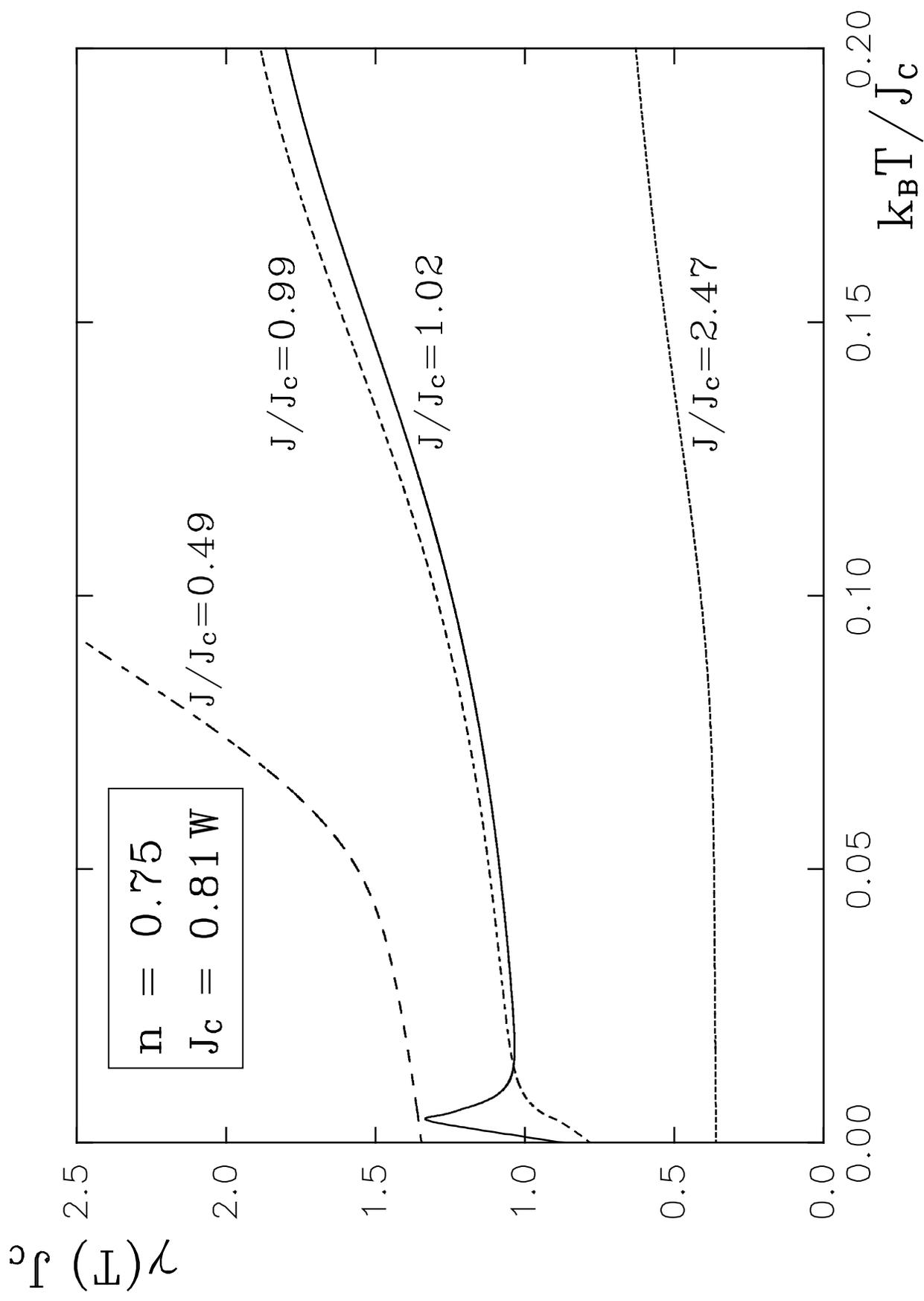
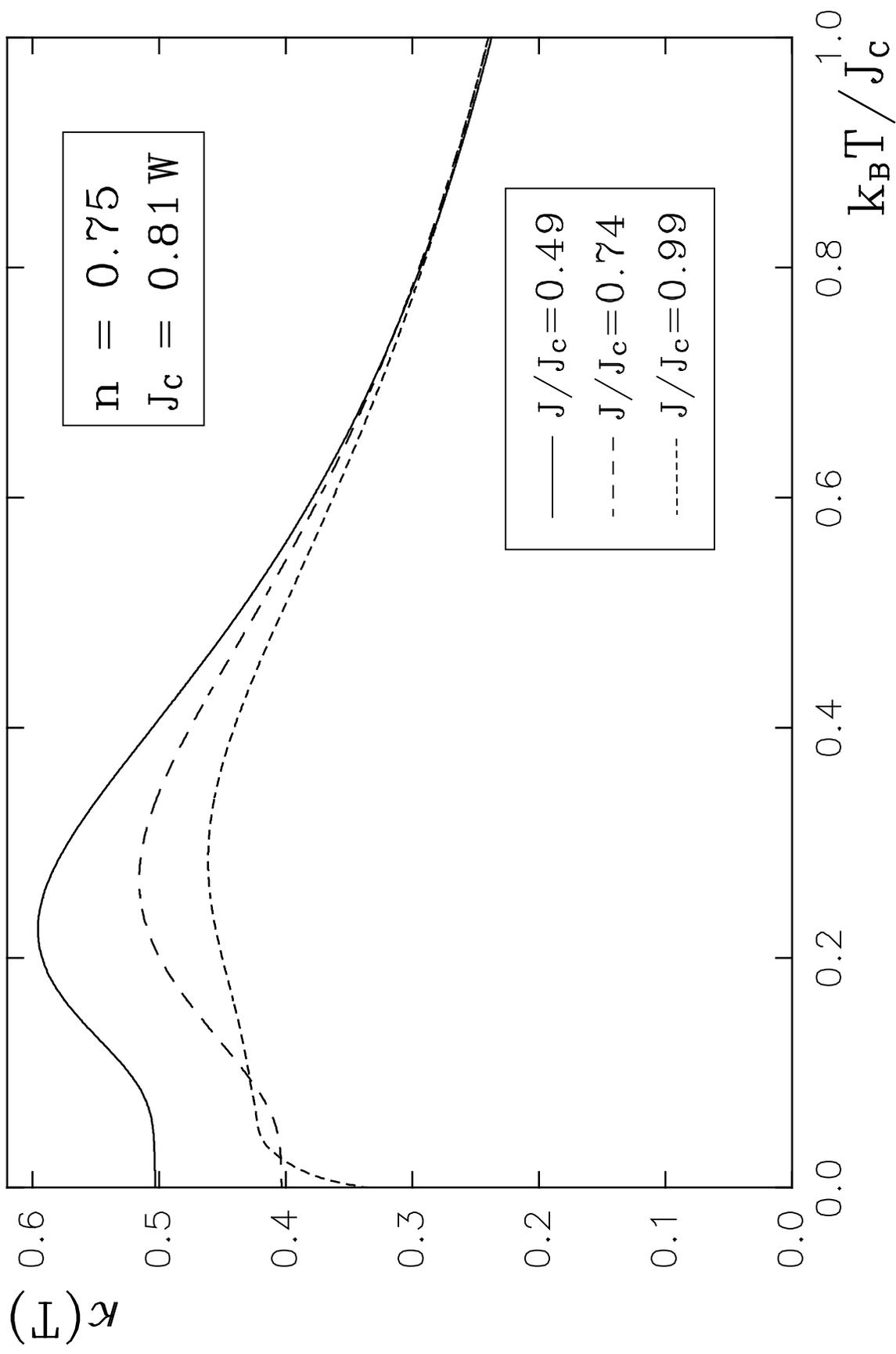
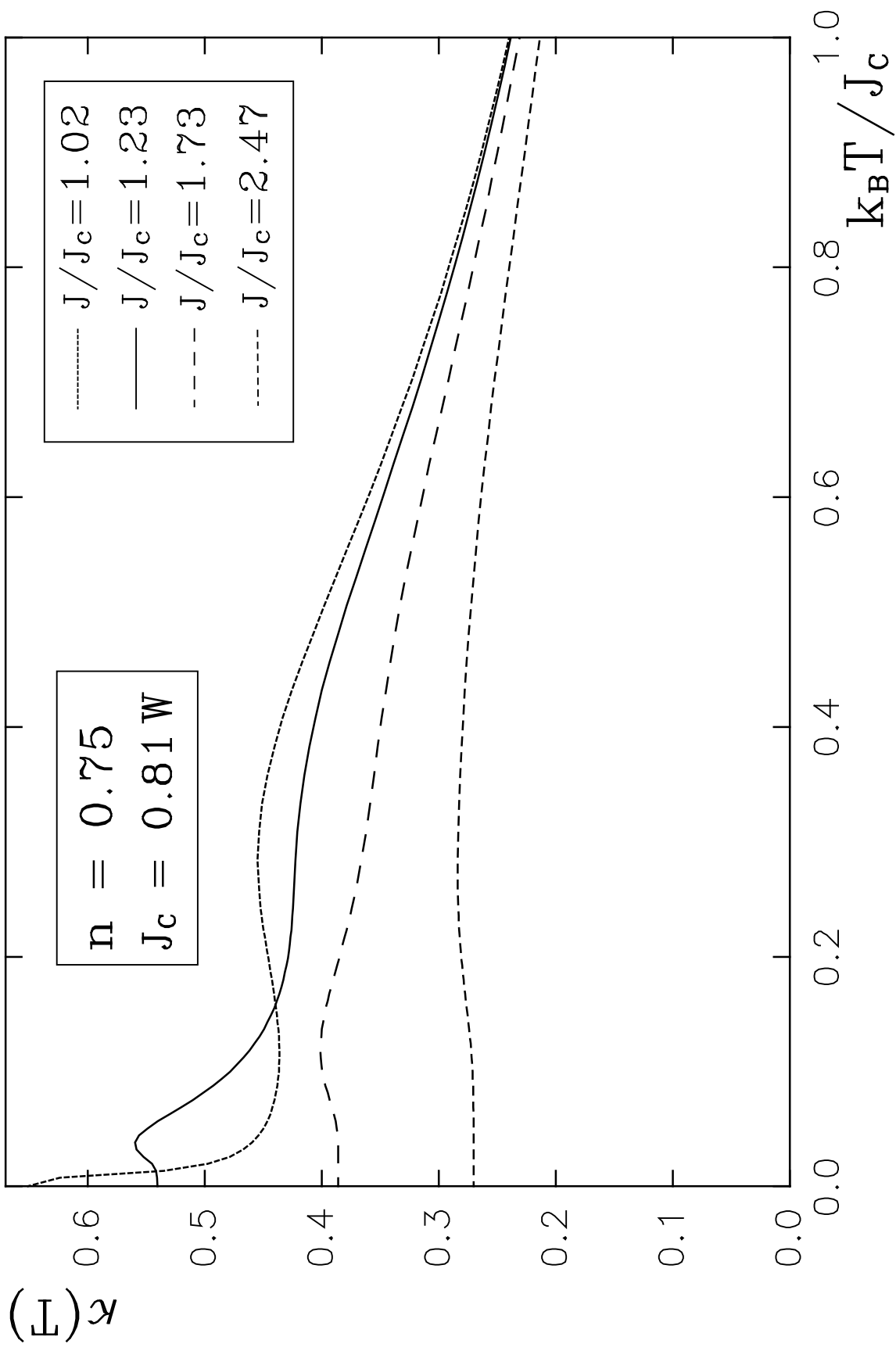
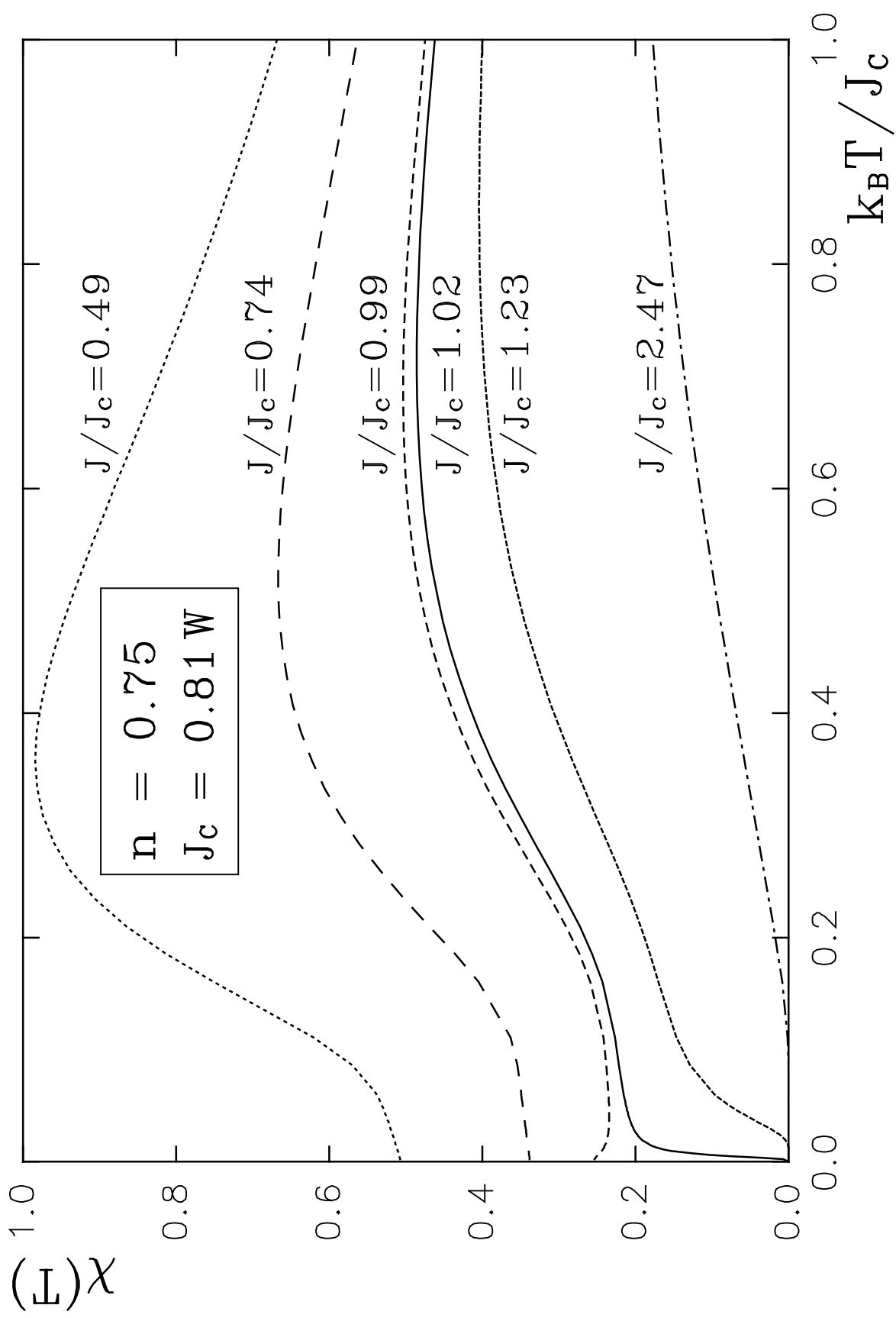
\end{document}